\documentclass[3p,fleqn,a4paper]{elsarticle}

\usepackage{graphicx}
\usepackage{wrapfig}
\usepackage{float}
\usepackage{epsfig}
\usepackage{tikz}
\usepackage{multirow}
\usepackage{amsfonts,amsmath,amssymb}
\usepackage[hang]{subfigure}
\usepackage{textcomp}
\usepackage{empheq}

\usepackage[left]{lineno}
\usepackage{mathrsfs}
\usepackage{amsthm}
\usepackage{listings}

\usepackage{bbm}
\usepackage{xspace}
\usepackage{yfonts}
\biboptions{sort&compress}

\newcommand{\pf}{\mathfrak{d}}
\newcommand{\grad}{\nabla}

\newcommand{\dx}{\text{d}\textbf{x} }

\newcommand{\BO}{\mathcal{B}}
\newcommand{\stress}{\boldsymbol{\sigma}}
\newcommand{\stresss}{{\sigma}}
\newcommand{\strain}{\boldsymbol{\varepsilon}}
\newcommand{\strainn}{{\varepsilon}}

\newcommand{\ML}{\mathcal{L}}
\newcommand{\MG}{\mathcal{G}}

\usepackage{mathrsfs}
\newcommand{\RN}[1]{%
  \textup{\uppercase\expandafter{\romannumeral#1}}%
}
\journal{---------}

\usepackage{pdfpages}
\usepackage[colorlinks=true,linkcolor=black, citecolor=blue, urlcolor=blue]{hyperref}

\usepackage{natbib}

\begin{document}


\begin{frontmatter}

\title{A Multi-Phase-Field Model for Fiber-Reinforced Composite Laminates Based on Puck Failure Theory}

\author[TUW] {Pavan Kumar Asur Vijaya Kumar}

\author[TUW]{Rafael Fleischhacker}

\author[ISD,SUST]{Aamir Dean \corref{cor2}}
\cortext[cor2]{Corresponding author}
\ead{a.dean@isd.uni-hannover.de}

\author[TUW]{Heinz E Pettermann}


\address[TUW]{Institute of Lightweight Design and Structural Biomechanics, Technische Universität Wien, Getreidemarkt 9, 1060 Vienna, Austria.}
\address[ISD]{Institute of Structural Analysis. Leibniz Universit\"at  Hannover, Appelstr. 9A, 30167 Hannover, Germany}
\address[SUST]{School of Civil Engineering, College of Engineering, Sudan University of Science and Technology, P.O. Box 72, Khartoum, Sudan}

\begin{abstract}

This article proposes a multi-phase-field model using the Puck failure theory to predict the failure in fiber-reinforced composites (FRCs) laminates. Specifically, this work proposes a two-dimensional multi-field model in conjunction with a mesh overlay method to compute in-plane damage in the FRCs laminates with various ply orientations. The formulation considers the two independent phase-field variables to trigger fiber and inter-fiber-dominated failure separately, thereby accessing the interrelation between the damage. Furthermore, the model considers two characteristic length scales and two structural tensors to describe the damage modes accurately. Each ply in the laminate is represented using a separate mesh and is combined using the mesh overlay method. Four benchmark examples are utilized to demonstrate the predictive capability of the proposed model. Specifically, coupon tests in tensile and compressive loading, open-hole tension, compact tension, and double-edged notched tension examples are presented along with the comparison with the experimental results from the literature. Furthermore, results regarding cross-ply laminates and isotropic laminates show the model's ability to mimic the experimental results both qualitatively and quantitatively.

\end{abstract}
\begin{keyword}
A. Multi-Phase-Field, B. Laminates,  C. Phase-field method, D. FRCs,  E. Fracture 
\end{keyword}
\end{frontmatter}



\section{Introduction}
\label{Introduction}

Laminated  composites (FRCs) have been the popular choice of engineering applications in the last few decades. This is due to their unique combination of mechanical, thermal, and chemical properties and, more importantly, the ability to manipulate the mechanical properties according to the designer's need. A prime example is the Boeing 787, with 50\%  of its weight and 80\%  by volume, and the Airbus A350 wide-body aircraft, with 53\%  of its weight being composites. The laminated composites have made their way into every lightweight alternative sector, such as aerospace, automotive, marine, energy, and sporting, to name a few. 

Despite their contributions, the FRCs possess a significant challenge in predicting fracture due to their complex failure mechanism \cite{BUI2021107705}. Numerical methods to predict the onset and damage propagation in the composites are highly desirable to supplement the experimental investigation and to enhance damage-tolerant designs. For a given loading, the propagation of cracks and the failure mode within the laminates highly depend on the layup sequence. Failure modes that can arise under in-plane loading include: 1. transverse intralaminar matrix cracking, 2. longitudinal intralaminar matrix cracking, 3. interlaminar cracking (delamination), 4. transverse crack propagation leading to fiber tensile or compressive failure, and 5. delamination growth \cite{Timon_review, CHENG2024110120}.

The literature on composites, especially FRCs, is dominated by experimental investigations. This article highlights the most relevant research. See \cite{Pinho2a} and the references for a more extensive review. Experimental investigation regarding the open-hole tension (OHT) of the laminates are presented in \cite{KAUL2021100097,FLATSCHER20121090,GREEN2007867_Heinz5,Hallett2009_Heinz6}. Specifically, \cite{CHUA_2} investigates the strength of the open hole tension and the underlying failure mode. \cite{GREEN2007867_Heinz5,CAMANHO_22,WISNOM2_23,HALLETT2009613,Hallett2009_Heinz6,Wisnom_26} considers various laminate sequences and the ply block to study the effects of stacking sequence, ply block thickness, and the effects of size in the laminates, among other factors. See \cite{Hallett2009_Heinz6,CHUA_2} for more details.  

Apart from the OHT, compact tension (CT) configurations are widely used to study laminates. The compact tension results for the carbon/epoxy are tested in \cite{Harris_R22,John_R23,underwood_R24,PINHO_R26,LAFFAN_R27,LAFFAN2_R28,LAFFAN_R29}, while  carbon/ carbon laminates are studied in \cite{DASSIOS_R31}, carbon/PEEK in \cite{REBER_R32}, boron/aluminum in \cite{Karl_R33,Sun_R34}. These studies also consider various laminate sequences under tension and compression. The notched tension specimens are extensively investigated to study their mechanical properties, including fracture toughness, and the effects of the notch in different structural configurations in various laminate sequences in \cite{Harris_R22,Karl_R33,morris_R45,Yeow_R46,underwood_R47,poe_R48,Evren_R51,Hallet_R52}, among others.

Several numerical approaches are used to model composites mathematically. In the related literature, various models are used to simulate damage propagation in FRCs, such as cohesive elements \cite{PINHO_R1}, smeared crack models \cite{PINHO_R2}, extended-finite element method (XFEM) \cite{MOES_R3}, peridynamics \cite{Bobaru_BR347,Xu_BR354}.  Additionally, a combined FEM and peridynamics approach \cite{OTERKUS_BR404}, discrete lattice models  \cite{BOYINA_BR404,MAYYA_BR426,BRAUN_BR442}  are also used. See \cite{CHENG2024110120} for an elaborate discussion on computational modeling of composite material, and \cite{Timon_review} for a review on general computational damage. The dominant approach stems from continuum damage mechanics (CDM), where unidirectional (UD) plies are treated as smeared (homogenized) materials within FEM. Various constitutive models for ply behavior are proposed, ranging from progressive degradation models \cite{Puck2002} to advanced continuum damage mechanics models \cite{allix_9_1,ladeveze_10_1,ladeveze_11_1}. Composites are modeled as anisotropic elastic-brittle continua in \cite{MATZENMILLER_15_1,LAPCZYK_16_1,MAIMI_17_1} or shear-driven inelasticity \cite{VANPAEPEGEM18_1}. Models addressing stiffness degradation and unrecoverable strain accumulation are proposed in \cite{barbero19_1,SCHUECKER21_1,SCHUECKER22_1,Schuecker2008}. The micro-mechanics inspired CDM model that includes nonlinearities stemming from the accumulation of unrecoverable strains, strain hardening, strain softening, and plasticity is proposed in \cite{KAUL2021100097,heinz2,FLATSCHER20121090,SCHUECKER22_1}. 


Phase-field (PF) methods, in general, are seen as a competition between the strain energy in the system and the energy required to create a new surface \cite{Griffith,Francfort1998,Bourdin2000,Bourdin2011}. Due to the smeared nature of the crack, the non-local approach, and the ease of integrating with different material models, the phase-field methods have become an appealing computational tool for fracture modeling. Consequently, phase-field methods are used to study fracture in brittle  \cite{Miehe,Schluter2014,MSEKH2015472,MIEHE20102765,MESGARNEJAD2015420,FREDDI2017156,PAGGI2017145,WU201772,NGUYEN2016567,CONTI20161033,BORDEN201277}, dynamic fracture \cite{BORDEN201277,Schluter2014,Bourdin2011,dynamic,ZHOU2019169,NGUYEN20181000},  thermo-mechanical fracture \cite{MIEHE2015486,PavanAsur,ASURVIJAYAKUMAR2022115096}, and composites \cite{BLEYER2018213,DEAN2020102495,WANG2020113270,PavanFGM,ALESSI20179,REINOSO2017205,HIRSHIKESH2019424,SINGH2021107348,ZHANG2021107371,YUE2023117432,DEAN2026111588} among others. In the context of the composites, the phase-field methods are primarily restricted to the study of single-ply/unidirectional fiber-reinforced composites and occasionally simple laminate sequences such as cross-ply laminates, see \cite{gultekin201823,gultekin2016542,REINOSO2017205, adria,BLEYER2018213}. Specifically, the laminates with various sequences are treated in and out-of-plane setting to study the effects of the laminate sequence. As a consequence, only crack propagation in the thickness plane (out of plane) is studied as in \cite{ZHANG,Mrunmayee,PRANAVI,PavanFRC,PavanP2FRC,adria,Ishank,trisha}. According to the authors, no articles within the context of phase-field methods consider more than two layers of laminate sequence to find the in-plane crack propagation in the laminates. 

Many PF models relying on anisotropic energies \cite{Ishank,ZHANG} fail to capture compressive loading due to the energy split. Furthermore, extending these methods to study laminates with different ply sequences is not trivial. Additionally, the models often resort to balancing the length scales to match the force-displacement curve. This approach is a direct consequence of the recent paper by Oscar et al. \cite{nucleationpaper,nucleation2}, which suggests that classical variational phase-field models cannot predict crack nucleation accurately. Furthermore, the study suggests that the entire failure initiation surface has to be included in the formulation for the phase-field model to describe crack nucleation. In other words, the crack nucleation has to be primarily accompanied by ad hoc criteria apart from the interpolation of the length scale $\ell $ (of PF). Within the context of the FRCs, the phenomenological criteria such as PUCK \cite{Puck2002}, Hashin \cite{Hashin}, or physically based criteria such as  \cite{Pinho1,Pinho2,Pinho4,Pinho5,Pinho1a,Pinho2a,Pinho3a}  have to be included in the model to account for the crack nucleation. See \cite{Revisiting} and the references therein for more extensive review of FRCs using Phase-field method.

This paper addresses the issues regarding the failure initiation (strength) surface by invoking the Puck failure theory. Furthermore, the in-plane damage in the laminates with various ply sequences is captured using a simple two-dimensional model employing mesh overlay techniques. As a consequence of the two-dimensional model, the study of the delamination is omitted. Further advantages and disadvantages of the model are detailed in the sequel. The article also compares the numerical results with the experimental methods in each example using crack propagation and qualitative assessment such as force vs. displacement and laminate mean stress, wherever available. 

The article is organized as follows. Section \ref{variational} presents the variation formulation using a multi-phase-field model. Section  \ref{puck} briefly introduces the Puck failure theory, followed by its integration into the variational framework. Thermodynamic consistency and the details of driving forces are presented in Section \ref{thermodynamic}. Section \ref{FEM} presents the finite element formulation in general for the n-ply in the laminates using mesh overlay techniques. The relevant information for the computer implementation of the mesh overlay technique is also presented. Section \ref{numex} considers four distinct geometries with various laminate sequences to present the model's predictive capacity. Specifically, the following examples, along with the comparison between the experiments and numerical methods, are presented. Section \ref{couponlaminates} presents the coupon test.  Section \ref{OHT} presents the results regarding open hole tension. Section \ref{compacttens_lam} presents the results regarding compact tension, and finally Section \ref{DNT} presents the results regarding the double edge notched tension. Finally, the conclusions are presented in Section \ref{conlusions}. The research data, along with the codes, are provided in Section \ref{data}.

\section{Variational Formulation}
\label{variational}

\begin{figure}[H]
    \centering
    \includegraphics[width=0.65\linewidth]{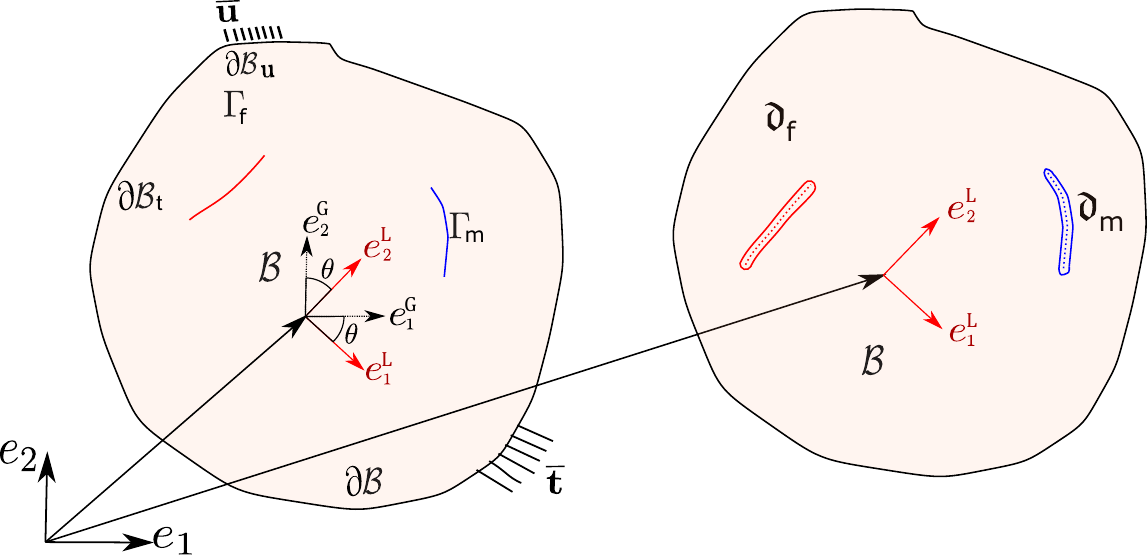}
    \caption{Geometrical description of the body under consideration. }
    \label{Figure1}
\end{figure}

This section proposes a multi-phase-field model for the fracture analysis of fiber-reinforced composite laminates. The damage at the ply level is initiated using Puck failure criteria, while displacement coupling stemming from laminates is used to determine the damage at the laminate level. Restricting the analysis to a two-dimensional setting, consider an arbitrary body $\mathcal{B} $ in an Euclidean space with its delimiting boundaries $\partial \mathcal{B}$. Then, for every position vector $\mathbf{x} \in \mathcal{B}$, vector valued displacement is defined as $\mathbf{u}(\mathbf{x}, t): \mathcal{B} \times [0,t] \rightarrow\mathbb{R}^{2}$, with pseudo time $\tau \in [0,t]$. The strain in the system is defined as a symmetric gradient of the displacement field $\boldsymbol{\varepsilon}(\mathbf{u}) = \grad^s\mathbf{u}$, with $\boldsymbol{\varepsilon}: \mathcal{B}  \rightarrow\mathbb{R}^{2 \times 2}  $. Prescribed boundary displacements are given by $\mathbf{u}=\mathbf{\bar{u}}\,\text{on}\,\partial\mathcal{B}_{\mathbf{u}}$, while the prescribed tractions are given by  $\bar{\mathbf{t}}=\boldsymbol{\sigma}\cdot\mathbf{n}$ applied on $\partial \mathcal{B}_{t}$. Here $\boldsymbol{\sigma}$ is the Cauchy stress tensor, and $ \mathbf{n}$ is the unit outward normal vector at the traction boundary. The boundary conditions satisfy  $\overline{\partial \mathcal{B}_{\mathbf{u}} \cup \partial \mathcal{B}_{t}} = \partial \mathcal{B}$ and $\partial \mathcal{B}_{\mathbf{u}} \cap \partial \mathcal{B}_{t} = \emptyset$. Furthermore, let $ \Gamma_f$ and $ \Gamma_m$ be crack sets for fiber failure and matrix failure, respectively, such that $ \Gamma_f, \Gamma_m   \subset \mathbb{R}^{1}$, and $ \Gamma_f\cup \Gamma_m=\Gamma$ as in Fig. \ref{Figure1}.

Griffith´s theory suggests that cracks are formed due to competition between the stored strain energy in the body and the energy required to create new surfaces. The strain energy in FRCs can be additively decomposed into energy associated with the fiber  $\Psi_f$ and the energy associated with the inter-fiber/matrix $\Psi_m$. Consequently, the energy required to create a new surface (herein referred to as surface energy) can be decomposed into fiber and inter-fiber surface energy. 

The phase-field methods to fracture approximate the sharp, discontinuous crack sets $ \Gamma$  using a diffusive crack model, represented by a scalar-valued field $\pf: \BO \times [0,t]\rightarrow [0,1]$, and the characteristic length scale $ \ell$. Here, $ \pf=0$ refers to the intact material, while $ \pf=1$ refers to the fully fractured material. Due to the presence of two distinct crack sets $ \Gamma_f$, and $ \Gamma_m  $, two independent phase-fields are introduced $ \pf_f$ and  $\pf_m$ representing a diffusive crack in fiber and inter-fiber, respectively.  

The total energy of the system takes the form 

\begin{equation}
      \Psi(\mathbf{u}, \Gamma_f, \Gamma_m )= \int_{\mathcal{B}\backslash\Gamma_f}  \Psi_f(\strain) \dx+ \int_{\mathcal{B}\backslash\Gamma_m}\Psi_m(\strain)\dx+  \int_{\Gamma_f} G_{C,f}\text{d}S +\int_{\Gamma_m} G_{C,m}\text{d}S-   \Psi_{ext}(\mathbf{u}).
      \label{total_energy}
\end{equation}

Here $\Psi_i(\strain)$ refers to the undamaged strain energy densities, while $G_{C, i}$ denotes the fracture energy densities corresponding to each $ (i=f,m)$. Herein, $(f)$ represents the fiber, and $(m)$ represents the inter-fiber. Furthermore, $ \Psi_{ext}(\mathbf{u})$ refers to the applied loading defined as

\begin{equation}
   \Psi_{ext}(\mathbf{u}) :=\int_{\mathcal{B}}\boldsymbol{f_{v}}\cdot \mathbf{u}\,\dx+\int_{\partial\mathcal{B}_t}\bar{\mathbf{t}}\cdot \mathbf{u}\,\text{dS}.
    \label{external_energy}
\end{equation}

Within the context of the phase-field approach, the surface energy can be approximated using the diffusive field as 

\begin{equation}
    \int_{\Gamma_i}G_{C,i}\text{ dS}\approx\int_{\mathcal{B}}\frac{G_{C,i}}{4c_{w}} 
\gamma_i(\pf_i, \grad \pf_i) \dx, 
\end{equation}

where $\gamma_i(\pf_i, \grad \pf_i)$ for each $(i=f,m)$ are crack surface energy density function given as 

\begin{equation}
    \gamma_i(\pf_i, \grad \pf_i)= \frac{\alpha(\mathfrak{d}_i)}{\ell_i}+\ell_i  \left[ \nabla\mathfrak{d}_i \cdot \mathcal{A}_i \cdot  \nabla\mathfrak{d}_i\right]. \quad  \quad (i=f,m)
    \label{surfaceenergy}
\end{equation}

Here $\alpha(\mathfrak{d}_i)$ is the geometric crack function such that $\alpha(0)=0 $, and $\alpha(1)=1 $ with $\alpha: [0,1]\rightarrow [0,1]$. The geometric crack function  $\alpha(\mathfrak{d}_i)$ is chosen in line with the AT2 model. i.e $\alpha(\mathfrak{d}_i)=\pf^{2}_i$. Furthermore, $\ell_i$ is the characteristic length scale, and $ \mathcal{A}_i$ is the second-order structural tensor to penalize the gradient of the phase-field. The structural tensor is written as 

\begin{equation}
    \mathcal{A}_f= e_2 \otimes e_2, \quad \quad  \mathcal{A}_m= e_1 \otimes e_1,
    \label{structuralTensor}
\end{equation}
where $e_1$ is the principal material direction along the fiber orientation, while  $e_2$ is the transverse in-plane orientation perpendicular to $e_1$. For the fiber orientation of $ \theta$ from the global X-coordinate, the material directions take the form

\begin{equation}
    e_1 =\begin{bmatrix}
\cos{\theta} \\
\sin{\theta}  
\end{bmatrix}, \quad \text{and } \quad   e_2 =\begin{bmatrix}
\sin{\theta} \\
\cos{\theta}  
\end{bmatrix}.\  
\end{equation}

The sequel introduces global and local ply coordinate systems to maintain consistency with the Puck failure criteria, which is based on the tri-axial stress state in local ply coordinates. Global quantities are denoted using  $(\cdot)^\MG$, while the local (ply) quantities are denoted using $(\cdot)^\ML$. The transformation from the local to global coordinates is achieved using the transformation matrix  $\mathcal{R}$ in line with \cite{LI2022107147,PAN2020112580,PAN2021113754,PAN2022114192} given as

 \begin{equation}
     \mathcal{R}=\begin{bmatrix}
         \cos{\theta}^2 & \sin{\theta}^2   & 2\cos{\theta}\sin{\theta} \\
         \sin{\theta}^2 & \cos{\theta}^2   & -2\cos{\theta}\sin{\theta}\\
       -  \cos{\theta}\sin{\theta} & \cos{\theta}\sin{\theta}   & \cos{\theta}^2-\sin{\theta}^2 
     \end{bmatrix}
     \label{Rotation}
 \end{equation}
The transformation of the strain field in the local ply setting is written as 
\begin{equation}
    \strain^{\ML}=\left( \mathcal{R}^{-1} \right)^{T} \cdot \strain^{\MG},
    \label{localstrain}
\end{equation}
while the stress field in local ply setting takes the form 

\begin{equation}
    \stress^{\ML}= \mathcal{R} \cdot \stress^{\MG}.
    \label{stress_local}
\end{equation}
The stress in the global system is then written as
\begin{equation}
    \stress^{\MG}=\mathbb{C}^{\MG} \cdot \strain^{\MG}.
    \label{stress_global}
\end{equation}
Furthermore, the constitutive stiffness matrix $ \mathbb{C}^\ML$  in the global setting can be derived from the local setting as 
\begin{equation}
       \mathbb{C}^{\MG}=\mathcal{R}^{-1} \cdot   \mathbb{C}^{\ML} \cdot \left( \mathcal{R}^{-1} \right)^{T}.
       \label{global_C}
\end{equation}

The local anisotropic constitutive stiffness matrix can be defined as a additive decomposition of the fiber and inter-fiber constituents as 

\begin{equation}
    \mathbb{C}^{\ML}=\mathbb{C}^{\ML}_f+\mathbb{C}^{\ML}_m,
\end{equation}
where $\mathbb{C}^{\ML}_f$ and $\mathbb{C}^{\ML}_m$ are the fiber and inter-fiber contribution whose specific description takes the form 

\begin{equation}
    \mathbb{C}^{\ML}_f=\begin{bmatrix}
        C_{11} & 0 & 0 \\
        0 & 0 & 0 \\
        0 & 0 & 0 
    \end{bmatrix}; \quad  \quad  \mathbb{C}^{\ML}_m=  \begin{bmatrix}
      0 &  C_{12}& 0 \\
          C_{12} &   C_{22}& 0 \\
        0& 0&   C_{33} 
    \end{bmatrix},
\end{equation}

with $C_{11} =\dfrac{E_{11}}{1-\nu_{12}\nu_{21}}$,$C_{12} =\dfrac{\nu_{12}E_{11}}{1-\nu_{12}\nu_{21}}$, $C_{22} =\dfrac{E_{22}}{1-\nu_{12}\nu_{21}}$,   and $C_{33} =G_{12}$. Furthermore $ E_{11}$, $ E_{22}$, and $ g_{12}$ are the Young's modulus, while $\nu_{12} $, and $\nu_{21}=\nu_{12} \dfrac{E_{11}}{E_{22}}$ are the Poisons´s ratio. 

With this at hand, the strain energy density contributions corresponding to the fiber and inter-fiber phases, respectively, are given by

\begin{equation}
    \Psi_f(\mathbf{u}, \pf)= g(\pf) \dfrac{1}{2} \strain^{\ML}\cdot \mathbb{C}^{\ML}_f\cdot \strain^{\ML}, 
    \label{fiberenergy}
\end{equation}

\begin{equation}
    \Psi_m(\mathbf{u}, \pf)= g(\pf) \dfrac{1}{2} \strain^{\ML}\cdot \mathbb{C}^{\ML}_m\cdot \strain^{\ML}.
    \label{matrixenergy}
\end{equation}

Due to the damage in the fiber and inter-fiber, the strain energy density contributions are degraded based on the degradation function $g(\pf)$ in line with the AT2 model. 
The total energy of the system considering the strain energy in Eqns. \eqref{fiberenergy}, \eqref{matrixenergy}, and surface energy  in Eqn. \eqref{surfaceenergy} takes the form 

\begin{eqnarray}
      \Psi(\mathbf{u}, \pf_f, \pf_m )= & \int_{\mathcal{B}} g(\pf) \dfrac{1}{2} \strain^{\ML}\cdot \mathbb{C}^{\ML}_f\cdot \strain^{\ML} \dx+ \int_{\mathcal{B}} g(\pf) \dfrac{1}{2} \strain^{\ML}\cdot \mathbb{C}^{\ML}_m\cdot \strain^{\ML} \dx  -   \Psi_{ext}(\mathbf{u}) \\ 
      +  & \int_{\mathcal{B}}\left[ G_{C,f}\frac{\alpha(\mathfrak{d}_f)}{\ell_f}+\ell_f  \left[\nabla\mathfrak{d}_f \cdot \mathcal{A}_f \cdot  \nabla\mathfrak{d}_f\right] \right]\dx+\int_{\mathcal{B}} \left[ G_{C,m}\frac{\alpha(\mathfrak{d}_m)}{\ell_m}+\ell_m \left[\nabla\mathfrak{d}_m \cdot \mathcal{A}_m \cdot  \nabla\mathfrak{d}_m\right] \right] \dx.
      \label{total_energy2}
\end{eqnarray}

Furthermore, due to multiple damage mechanisms, the degraded constitutive matrix in the local co-ordinates takes the form 

\begin{equation}
    \tilde{\mathbb{C}}^{\ML} =g(\pf)\mathbb{C}^{\ML}_f+g(\pf)\mathbb{C}^{\ML}_m,
\end{equation}

with 

\begin{equation}
    g(\pf)=\left\{ \begin{array}{c}
g_{f}\\
g_{m}\\
\text{min}\{g_{f},g_{m}\}
\end{array}\right\}
\label{degradedC}.
\end{equation}

Here $g_f=\left( 1-\pf_f\right)^2 $, and $g_m=\left( 1-\pf_m\right)^2 $  are the fiber and inter-fiber degradation functions respectively in line with standard AT2 formulation \cite{Bourdin2011}. Consequently, the degraded stress in the local co-ordinates takes the form 

\begin{equation}
    \tilde{\stress}^{\ML}= g(\pf)^T \stress^{\ML};  \quad \quad   \tilde{\stress}^{\MG}= \mathcal{R}^{-1} \cdot \tilde{\stress}^{\ML}. 
    \label{degradesstress}
\end{equation}
The initiation of the damage relies on the Puck failure theory based on the local stress state as detailed in the next section. 

\section{PUCK Failure Criteria}
\label{puck}

Puck failure theory \cite{Puck2002} distinguishes between the damage in fiber and inter-fiber independent of each other. The theory identifies two modes for fiber failure, namely: (i) fiber failure in tension, (ii) fiber failure in compression, and three modes of failure in the inter-fiber failure, namely: (iii) Mode-A (transverse tension or transverse tensile-shear failure), (iv) Mode-B (transverse compression + in-plane shear), and (v) Mode-C (Dominant transverse compression with negligible shear). Each failure mode is triggered when the energetic exposure factor corresponding to the failure mode, denoted as $F_i $, reaches the value 1. Puck failure criteria in a two-dimensional setting in the local ply coordinates are presented in the sequel. Here, subscript 1 represents the principle material direction (fiber direction), while subscripts 2 and 3 represent the direction normal to the fiber and direction in-plane and out-of-plate, respectively.  


In the local ply coordinates, $E_{11}$, $E_{22}$, are the longitudinal modulus, while the shear modulus is denoted by $E_{12}$. Similarly, the longitudinal strengths in the fiber direction and transverse tensile strengths are denoted by $R_{11}^T$, $R_{22}^T$, and $R_{12}$ respectively. Furthermore, Puck failure theory allows for a distinction between tensile and compressive failure. Consequently, the longitudinal strength counterparts in compression are represented using $R_{11}^C$ and $R_{22}^C$. The major poisson's ratios are represented as $ \nu_{12}$, and $ \nu_{23}$, while the puck matrix inclination parameters are represented using  $P_{21}^+ $, $P_{21}^- $, $P_{22}^+ $, and $P_{22}^- $  with usual notations. See \cite{Puck2002,PavanFRC,Revisiting} for detailed description of the Puck failure theory. 


\subsection{Fiber Failure }

The fiber exposure factor in tension $F_{f}^T$ is defined as a ratio of the stress in the principle material direction ($\stresss^{\ML}_{11}$) to the strength of the fibers. i.e 

\begin{equation}
    F_{f}^T=\dfrac{\stresss^{\ML}_{11}}{R_{11}^T},    \quad  \text{if} \quad \stresss^{\ML}_{11}>0. 
\end{equation}


Similarly,  the fiber exposure factor in compression takes the form
\begin{equation}
    F_{f}^C=\sqrt{ \left( \dfrac{\stresss^{\ML}_{11}}{R_{11}^C}\right)^2 + \alpha\left( \dfrac{\stresss^{\ML}_{12}}{R_{12}}\right)^2 }    \quad  \text{if} \quad \stresss^{\ML}_{11}<0. 
\end{equation}
Here $\alpha \in [0,1]$ is the influence of the shear on the compressive fiber failure.

\subsection{Matrix Failure}

If stress $ \stresss^{\ML}_{22} >0$, the exposure factor for the inter-fiber failure in  Mode-A transverse tension takes the form 

\begin{equation}
    F_m^T= \sqrt{  \left( \dfrac{\stresss^{\ML}_{12}}{R_{12}}\right)^2 + \left( 1-P_{21}^+  \dfrac{R_{22}^T}{R_{12}} \right)^2 + \left( \dfrac{\stresss^{\ML}_{22}}{R_{22}^T}\right)^2    } + \dfrac{P_{21}^+ \stresss^{\ML}_{22} }{R_{12}}.
\end{equation}

The Mode-B of Puck criteria refers to inter-fiber dominated transverse compression. The exposure factor for Mode-B is triggered, if the local stress state fulfills   $ \stresss^{\ML}_{22} <0$ as well as 
\begin{equation}
    \left[   \left| \dfrac{\stresss^{\ML}_{22}}{\stresss^{\ML}_{12}}  \right| >0  \right]  \& \&   \left[   \left| \dfrac{\stresss^{\ML}_{22}}{\stresss^{\ML}_{12}}  \right| <  \left| \dfrac{R_{22}^a}{R_{12}^c}  \right|  \right],
\end{equation}
where $ R_{22}^a =   \dfrac{R_{12}}{2 P_{21}^-} \left\{  \sqrt{1+ 2 P_{21}^- \dfrac{R_{22}^C}{R_{12}}}  -1 \right\} $, and $ R_{12}^c = R_{12} \sqrt{1+2P_{22}^- }$.
The exposure factor for the compression then takes the form
\begin{equation}
    F_m^C= \sqrt{  \left( \dfrac{\stresss^{\ML}_{12}}{R_{12}}\right)^2 + \left( P_{21}^- \dfrac{\stresss^{\ML}_{22}}{R_{12}} \right)^2  } + \dfrac{P_{21}^- \stresss^{\ML}_{22} }{R_{12}}. 
\end{equation}
Similarly, if the stress in local ply coordinates
\begin{equation}
    \left[   \left| \dfrac{\stresss^{\ML}_{12}}{\stresss^{\ML}_{22}}  \right| >0  \right]  \& \&   \left[   \left| \dfrac{\stresss^{\ML}_{12}}{\stresss^{\ML}_{22}}  \right| <  \left| \dfrac{R_{12}^c}{R_{22}^a}  \right|  \right]
\end{equation}
is applicable, then the compression exposure factor corresponding to the matrix failure in Mode-C corresponding to out-of-plane transverse compression can be written as
\begin{equation}
    F_m^C= \left(\dfrac{\stresss^{\ML}_{12}}{2 (1+P_{22}^-)R_{12}} \right)^2 -\left(\dfrac{\stresss^{\ML}_{22}}{R_{22}^C} \right)^2 \dfrac{R_{22}^C}{\stresss_{22}^{\ML}}.
\end{equation}

The energy corresponding to the local stress state  $\stress^{\ML}_* $ when the Puck failure criteria are met (for one or multiple failure modes) can be computed for fiber ($\Psi_{f,0}$) and the inter-fiber ($\Psi_{m,0}$) as


\begin{equation}
\Psi_{f,0}= \begin{cases}
    \dfrac{1}{2} \stresss^{\ML}_{11} \strainn^{\ML}_{11}  & \text{if }  F_f^T,  F_f^C \geq 1  \\
0 & \text{if }  F_f^T,  F_f^C < 1
\end{cases}.
\label{fiberPuck}
\end{equation}
\begin{equation}
\Psi_{m,0}= \begin{cases}
    \dfrac{1}{2} \left( \stresss^{\ML}_{22} \strainn^{\ML}_{22} +2\stresss^{\ML}_{12} \strain^{\ML}_{12}\right)  & \text{if }  F_m^T,  F_m^C \geq 1  \\
0 & \text{if }  F_m^T,  F_m^C < 1
\end{cases}.
\label{matrixPuck}
\end{equation}

\section{ Thermodynamic Consistency and Driving Force}
\label{thermodynamic}

The second law of thermodynamics that ensures the consistency of the variational formulation is presented using the triplet $ (\mathbf{u}, \pf_f, \pf_m)$ as 

\begin{equation}
      \mathcal{D} =\left( \stress -\dfrac{\partial \Psi}{\partial \strain}\right) \cdot \dot{\strain}-  \left( \dfrac{\partial\Psi }{\partial \pf_f} \dot{\pf}_f +    \dfrac{\partial\Psi }{\partial \grad\pf_f} \grad\dot{\pf}_f\right) -  \left( \dfrac{\partial\Psi }{\partial \pf_m} \dot{\pf}_m +    \dfrac{\partial\Psi }{\partial \grad\pf_m} \grad\dot{\pf}_m\right) \geq 0. 
\end{equation}

which is referred to as classius-Duhem inequality. Since stress can be written as  $\stress =\dfrac{\partial \Psi}{\partial \strain}= \tilde{\mathbb{C}}^{\MG}:\strain^{\MG}$ in the global setting, the classius-Duhem inequality takes the form 

\begin{equation}
      \mathcal{D} = \left( \stress -\dfrac{\partial \Psi}{\partial \strain}\right) \dot{\strain}-  \mathcal{D}_f \dot{\pf}_f+\mathcal{D}_m \dot{\pf}_m \geq 0,
      \label{Doriginal}
\end{equation}
where $\mathcal{D}_f $, and $\mathcal{D}_m $ are crack energy associated with fiber and inter-fiber respectively, whose specific form is written as 

\begin{eqnarray}
    \mathcal{D}_f= 2(1-\pf_f) \Psi- G_{C,f} \left[ \dfrac{\pf_f}{\ell_f}- \ell_f \grad \pf_f \cdot\mathcal{A}_f \cdot \grad \pf_f  \right] \label{Df}, \\
        \mathcal{D}_m= g'(\pf) \Psi- G_{C,m} \left[ \dfrac{\pf_m}{\ell_m}- \ell_m \grad \pf_m \cdot\mathcal{A}_m \cdot \grad \pf_m  \right]. \label{Dm}
\end{eqnarray}
Note that $g'(\pf) \Psi$ is not defined due to $ \text{min}(g_f, g_m)$. Consequently, first minima is taken followed by the  $g'(\pf)$. Here the expression $ G_{C,i} \left[ \dfrac{\pf_i}{\ell_i}- \ell_i \grad \pf_i \cdot\mathcal{A}_i \cdot \grad \pf_i  \right] $ for each $ i=f,m$ refers to the energetic crack resistance. Furthermore, $g'(\pf) \Psi$ is the crack driving force.

A direct consequence of Eqn. \eqref{Doriginal} and the boundedness of the phase-field leads to the first-order optimality condition, also known as the Karush-Kuhn-Tucker (KKT) condition, whose specific form reads 

\begin{eqnarray}
    \dot{\pf}_f \geq 0, -\mathcal{D}_f \leq 0,  \mathcal{D}_f \dot{\pf}_f=0, \\
    \dot{\pf}_m \geq 0, -\mathcal{D}_m \leq 0,   \mathcal{D}_m  \dot{\pf}_m=0.
\end{eqnarray}

In order to enforce the irreversibility of the phase-field variables, the driving force based on the Puck failure theory takes the form

\begin{equation}
    \mathcal{H}_f = \max_{\tau \in [0,T]} \left\{ \zeta_f \left(\dfrac{\Psi_f-\Psi_{f, 0}}{G_{C,f}} \right) \right\}  \quad \text{if }\quad \Psi_{f, 0} >0, \text{and}  \quad 0 \quad \text{otherwise},
    \label{fiberDriving}
\end{equation}
\begin{equation}
    \mathcal{H}_m = \max_{\tau \in [0,T]} \left\{ \zeta_m \left(\dfrac{\Psi_m-\Psi_{m, 0}}{G_{C,m}}  \right) \right\}  \quad \text{if }\quad \Psi_{m, 0} >0, \text{and}  \quad 0 \quad \text{otherwise}. 
    \label{matrixDriving}
\end{equation}
Here  $ \zeta_f$ and $\zeta_m $ are the dimensionless driving parameters to characterize the propagation of the fiber and inter-fiber failure. 


 \section {Finite Element Implementation: Compact UEL}
\label{FEM}

The solution to the  Eqn. \ref{total_energy2} can be obtained by solving the total energy potential as a minimization problem with respect to triplet  $(\mathbf{u}, \pf_f, \pf_m) $ at every discrete time step $\tau \in [0,t]$, i.e to determine $(\mathbf{u}, \pf_f, \pf_m) $ from 
\begin{equation}
    (\mathbf{u}^*, \pf_f^*, \pf_m^*)= \text{Arg min}_{\mathcal{S}} \Psi(\mathbf{u}, \pf_f, \pf_m) 
\end{equation}

with  $\mathcal{S}=\{ \dot{\pf_f}, \dot{\pf_m} \geq 0, \text{ for all } \mathbf{x}\in\mathcal{B}\}$. For any admissible test function  $(\delta\mathbf{u}, \delta\pf_f,\delta \pf_m) \in (\mathfrak{A}_u, \mathfrak{A}_f, \mathfrak{A}_m) $, the first variation of the total energy functional leads to the following multi-phase-field weak form 

 \begin{eqnarray}
   \int_{\mathcal{B}} \tilde{\stress}^{\MG}: \strain^{\MG} (\delta \mathbf{u}) \dx  - \int_{\mathcal{B}}\boldsymbol{f_{v}}\cdot \delta\mathbf{u}\,\dx+\int_{\partial\mathcal{B}_t}\bar{\mathbf{t}}\cdot \delta\mathbf{u}\,\text{dS}=0, \\
    \int_{\mathcal{B}} G_{C,f} \left[ \dfrac{\pf_f}{\ell_f} \delta \pf_f+ \ell_f  \grad \pf_f \cdot \mathcal{A}_f \cdot \grad \delta \pf_f \right]\dx - \int_{\mathcal{B}} \left( \tilde{\stress}^{\MG}: \strain^{\MG} \delta \pf_f \right) \, \dx =0,\\
\int_{\mathcal{B}} G_{C,m} \left[ \dfrac{\pf_m}{\ell_m} \delta \pf_m+ \ell_m \grad \pf_m \cdot \mathcal{A}_m \cdot \grad \delta \pf_m \right]\dx - \int_{\mathcal{B}} \left(\tilde{\stress}^{\MG}: \strain^{\MG} \delta \pf_m \right) \, \dx =0. 
 \end{eqnarray}
 with triplet $(\mathbf{u}, \pf_f, \pf_m) \in (\mathfrak{B}_u, \mathfrak{B}_f, \mathfrak{B}_m) $. Recalling the standard Bubnov-Galerkin method, the functional space for the triplet $(\mathbf{u}, \pf_f, \pf_m)$ is given by 

 \begin{eqnarray}
    \mathbf{u} \in \mathfrak{B}_u: =\{ \mathbf{u} \in H^1(\mathcal{B})| \grad \mathbf{u}=\bar{u} \text{ on } \partial \mathcal{B}_u  \}, \\
        \pf_f \in \mathfrak{B}_f: =\{ \pf_f \in H^1(\mathcal{B})| \pf_f \in [0,1],   \dot{\pf}_f \geq 0,  \text{ for all } \mathbf{x} \in  \mathcal{B}  \}, \\
       \pf_m \in \mathfrak{B}_m: =\{ \pf_m \in H^1(\mathcal{B})| \pf_m \in [0,1],   \dot{\pf}_m \geq 0,  \text{ for all } \mathbf{x} \in  \mathcal{B}  \},
\end{eqnarray}
 
while the functional space for the variations $(\delta\mathbf{u}, \delta\pf_f,\delta \pf_m)$ takes the form 
   \begin{eqnarray}
   \delta \mathbf{u} \in \mathfrak{A}_u: =\{ \delta \mathbf{u} \in H^1(\mathcal{B})| \grad  \delta\mathbf{u}= 0 \text{ on } \partial \mathcal{B}_u  \}, \\
        \delta\pf_f \in \mathfrak{A}_f: =\{ \delta \pf_f \in H^1(\mathcal{B})| \grad \delta \pf_f \geq 0, \text{ for all } \mathbf{x} \in  \mathcal{B}  \}, \\
       \delta\pf_m \in \mathfrak{A}_m: =\{ \delta \pf_m \in H^1(\mathcal{B})| \grad \delta \pf_m  \geq 0   \text{ for all } \mathbf{x} \in  \mathcal{B}  \}.
\end{eqnarray}

Let the functional space  $ \mathcal{B}$  is discretized into $n_e$ non-overlapping isoparametric elements such that $\mathcal{B} \approx  \bigcup_{e=1} ^{n_{e}} \mathcal{B}^{(e)}$ and the partition of unity holds. Let $ \mathcal{B}^{(N_e)}$ is the nodes associated with each element $ \mathcal{B}^{(e)}$. Then, we can create a set of $i$ elements sharing the same node set  $ \mathcal{B}^{(N_e)}$. i.e 

\begin{equation}
     \mathcal{B}^{(e)}_{i} = \mathcal{B}^{(e)} \quad \text{ for \quad i=1,2,...,j }
\end{equation}
where $j$ is the number of plies in the laminates with each ply having an fiber orientation of $\theta_i $. Specifically, $\mathcal{B}^{(e)}$ is refered to as a base mesh, while the $ \mathcal{B}^{(e)}_{i}$ for each $i=1,2,...,j$ are referred to as dummy mesh. The dummy mesh also fulfills the following condition

\begin{equation}
    \mathcal{B} \approx  \bigcup_{e=1} ^{n_{e}} \mathcal{B}^{(e)} =  \bigcup_{e=1} ^{n_{e}} \mathcal{B}^{(e)}_i \quad \text{ for each \quad i=1,2,...,j }
\end{equation}
where the node set $ \mathcal{B}^{(N_e)}$ is shared among all the element sets $\mathcal{B}^{(e)}_i $.

\begin{figure}
    \centering
    \includegraphics[width=0.65\linewidth]{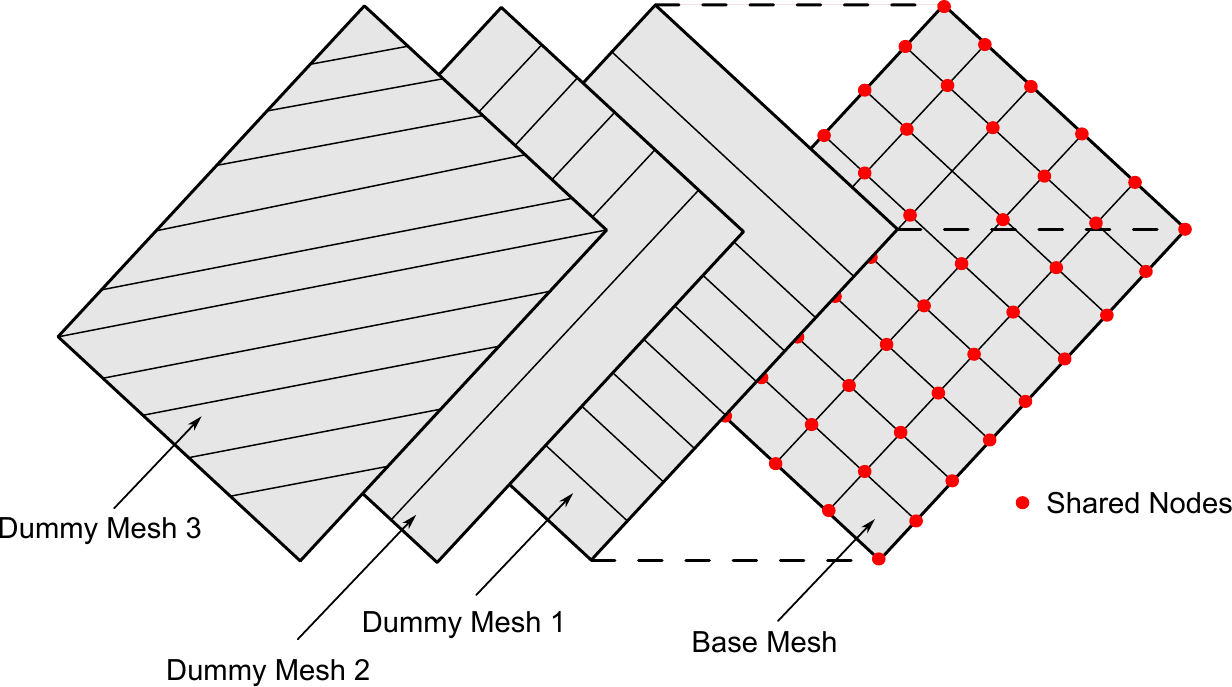}
    \caption{Geometric description of the mesh overlay method.}
    \label{meshoverlay}
\end{figure}

Fig. \ref{meshoverlay} describes this method as the mesh overlay method. The nodes are shared among all the layers, while each layer contains a unique element set with unique fiber orientation $ \theta_i$. Consequently, the displacement is coupled in all the layers, while the stresses are decoupled and depend on the fiber orientation $ \theta_i$ for each layer. Consequently, the energy developed in each layer is a function of the fiber orientation. 

With this setting, the primary fields and their respective gradients can be interpolated for each layer $i=1,2,...,j$ as 

\begin{equation}
\mathbf{u}^{e}=  \sum_{k=1}^{n_e} \textbf{N}_{k}^{u} \mathbf{u}_{k}^{e}, \; \;\;\;\;
\mathfrak{d}_{f}^{e}=  \sum_{k=1}^{n_e} \textbf{N}_{k}^{\mathfrak{d}} \mathfrak{d}_{f,k}^{e}, \; \;\;\;\; 
\mathfrak{d}_{m}^{e}=  \sum_{k=1}^{n_e} \textbf{N}_{k}^{\mathfrak{d}} \mathfrak{d}_{m,k}^{e}, \; \;\;\;\; 
\label{eq: Interpolation function}
\end{equation}
\begin{equation}
\grad \mathbf{u}^{e}=  \sum_{k=1}^{n_e} \textbf{B}_{k}^{u} \mathbf{u}_{k}^{e}, \; \;\;\;\;
\grad \mathfrak{d}_{f}^{e}=  \sum_{k=1}^{n_e} \textbf{B}_{k}^{\mathfrak{d}} \mathfrak{d}_{f,k}^{e}, \; \;\;\;\; 
\grad \mathfrak{d}_{m}^{e}=  \sum_{k=1}^{n_e} \textbf{B}_{k}^{\mathfrak{d}} \mathfrak{d}_{m,k}^{e}, \; \;\;\;\; 
\label{eq: Interpolation function}
\end{equation}

where $ \textbf{N}_k^{u}$, and $\textbf{N}_{k}^{\mathfrak{d}} $ are the shape functions associated with the node $k$ for the fields $ \mathbf{u}$, and two phase-fields $(\pf_f, \pf_m)$ respectively. Furthermore $ \textbf{B}_k^{u}$, and $ \textbf{B}_k^{\pf}$ are the derivatives of the shape functions at the node $k$ for displacement $ \mathbf{u}$ and the two-phase-fields, respectively. The triplet   $(\mathbf{u}^e, \pf_f^e, \pf_m^e)$ is computed using the shape functions derived from integration points at the nodal level leading to a coupling between displacement and the damage in the laminates. Consequently, stress and strains at individual ply can be determined, while the damage within each ply cannot be obtained through mesh overlay methods. Additionally, the exposure factor related to Puck modes at individual ply can still be computed.

With the isoparametric interpolation functions, the discrete version of the elemental residual vectors for each layer $ i=1,2,...,j$  takes the form
\begin{equation}
\mathbf{R}^{e}_{\mathbf{u},i}=\int_{ \mathcal{B}^{(e)}_i} (\textbf{B}^{u})^{T}  \tilde{\sigma }^{\MG} \dx - \int_{ \mathcal{B}^{(e)}_i} (\textbf{N}^{u})^{T} \mathbf{f}_{v}\dx- \int_{ \partial \textbf{B}^{(e)}_i}(\textbf{N}^{u})^{T} \mathbf{\overline{t}}\text{dS},
\label{Res-1}
\end{equation}
\begin{equation}
\mathbf{R}^{e}_{f,i}=\int_{ \mathcal{B}^{(e)}_i} -2(1-\pf_f) (\textbf{N}^{\pf})^T \mathcal{H}_f +\dfrac{G_{C,f}}{\ell_f}\left[ (\textbf{N}^{\pf})^T\pf_f +\ell_f^2 (\textbf{B}^{\pf})^T \mathcal{A}_f \grad \pf_f \right]\dx,
\label{Res-2}
\end{equation}
\begin{equation}
\mathbf{R}^{e}_{m,i}=\int_{ \mathcal{B}^{(e)}_i} -2(1-\pf_m) (\textbf{N}^{\pf})^T \mathcal{H}_m +\dfrac{G_{C,m}}{\ell_m}\left[ (\textbf{N}^{\pf})^T\pf_m +\ell_m^2 (\textbf{B}^{\pf})^T \mathcal{A}_m \grad \pf_m \right]\dx.
\label{Res-3}
\end{equation}
If the displacement and the tractions are applied on the nodes, the external load is shared among all the plies in the laminates.  The corresponding Newton-Raphson iteration for the globally assembled non-linear system at $(n+1)$ the step for each layer $ i=1,2,..j$ can be expressed as 
 \begin{equation}
    \left[\begin{array}{c}
\mathbf{u}\\
{\mathfrak{d}_{f}}\\
{\mathfrak{d}_{m}}
\end{array}\right]_{n+1}=\left[\begin{array}{c}
\mathbf{u}\\
{\mathfrak{d}_{f}}\\
{\mathfrak{d}_{m}}
\end{array}\right]_{n}-\left[\begin{array}{ccc}
K^{\mathbf{uu}} & 0 & 0\\
0 & K^{ff} & 0\\
0 & 0 &  K^{mm}
\end{array}\right]_{n+1}^{-1}\left[\begin{array}{c}
\mathbf{R_{u}}\\
\mathbf{R}_f\\
\mathbf{R}_m
\end{array}\right]_{n},
 \end{equation}

where the particular form of the elemental stiffness at the element level reads
\begin{equation}
\label{eps}
{K}^{\mathbf{uu}}_i := \dfrac{\partial \mathcal{R}^e_{\mathbf{u},i} }{\partial \mathbf{u}^{e}} =  \int_{ \mathcal{B}^e_i} (\textbf{B}^{u})^{T} \tilde{\mathbb{C}}_{\MG}\textbf{B}^{u} \dx,
\end{equation}
\begin{equation}
\label{eps2}
{K}^{ff}_i := \dfrac{\partial \mathcal{R}^e_{f,i} }{\partial \pf_f^{e}} =  \int_{ \mathcal{B}^e_i}  \left(2 \mathcal{H}_f + \dfrac{G_{C,f}}{\ell_f} \right)(\textbf{N}^{\pf})^T \textbf{N}^{\pf} +G_{C,f} \ell_f (\textbf{B}^{\pf})^T \mathcal{A}_f \textbf{B}^{\pf} \dx,
\end{equation}
\begin{equation}
\label{eps3}
{K}^{mm}_i := \dfrac{\partial \mathcal{R}^e_{m,i} }{\partial \pf_m^{e}} =  \int_{ \mathcal{B}^e_i}  \left(2 \mathcal{H}_m + \dfrac{G_{C,m}}{\ell_m} \right)(\textbf{N}^{\pf})^T \textbf{N}^{\pf} +G_{C,m} \ell_m (\textbf{B}^{\pf})^T \mathcal{A}_m \textbf{B}^{\pf} \dx.
\end{equation}

The above system of equations has been implemented in Abaqus-UEL to take advantage of the in-built Newton-Raphson solver, mesh assembly, and the automatic time-stepping scheme.  The corresponding codes are provided in the Data Availability section.

To this end, the present model is developed to predict the laminate response under the consideration of anisotropic stiffness degradation. The model does not incorporate plasticity, strain hardening, or softening behavior. Consequently, the material non-linearity is not addressed. Furthermore, the two-dimensional nature of the model impose certain limitations. The interfaces between the plies are assumed to be perfect. Consequently, delamination is not accounted for in the present model. Additionally, the model is not recommended when the out-of-the-plane stresses are expected to have a pronounced influence on the laminate behavior. Furthermore,  the influence of the stacking sequence cannot be captured due to the mesh overlay method, while ply thickness effects can be captured to some extent by scaling the load-displacement.

\section{Numerical Examples} \label{numex}

The formulation offers several novel aspects regarding the crack propagation in the laminates. The interconnection between the different failure mechanisms is captured well using the present model. Furthermore, comparisons with the experiments are made both qualitatively and quantitatively to show the predictive capabilities of the proposed model. Even though the mesh overlay method simplifies the model to exclude delamination, the method offers computationally efficient alternatives to the three-dimensional models. Four structures are considered for the comparison as listed below:

1. The coupon tests with uni-axial loading in compression and tension with different layup sequences are considered in line with the test conducted by Flatscher et al. \cite{Flatscherdiss}. Specifically,  laminates with a layup of $[(\theta/ -\theta)_8]_s$ with $\theta = 45^\circ,\ 60^\circ,\ 75^\circ$ are considered. The resulting transverse compression and tension stress state are compared with the experimental observation. 

2. The open hole tension subjected to uniaxial tension with layup sequence of $[0^\circ_4/90^\circ_4]_s$ and $[45^\circ_4/-45^\circ_4]_s$ are compared with the experimental results in line with \cite{FLATSCHER20121090}. The results compare the onset and interaction of the damage mechanism in the laminate.

3. A blocked compact tension with the layup sequence of $[0^\circ_2/90^\circ_2]_{2s}$, and $[0^\circ_4/90^\circ_4]_{2s}$ are compared both qualitatively and quantitatively against the experimental results from \cite{LI20091891}. 

4. Finally, the double-notched tension with three different layup sequences comprising of cross-ply and isotropic laminates are compared with the experiments from \cite{Hallet_R52} to show the predictive capability of the proposed model. The interaction between the fiber-dominated and inter-fiber-dominated failure is captured.

\subsection{Coupon Tests} \label{couponlaminates}

\begin{table}[h]
    \centering
    \begin{tabular}{|c|c|c|c|} \hline 
         Tests&  Layup Sequence & $L$ (mm) & $W$ (mm) \\ \hline 
         T-45 &  $[(45^\circ/ -45^\circ)_8]_s$ & $134$ & $32$ \\ 
         C-45 &  $[(45^\circ/ -45^\circ)_8]_s$ & $56$& $32$\\
         C-60 &  $[(60^\circ/ -60^\circ)_8]_s$ & $32$& $32$\\
         T-75, C-75 &  $[(75^\circ/ -75^\circ)_8]_s$ & $20$& $32$\\
         \hline
    \end{tabular}
    \caption{Length, width and layup sequences of the specimens in coupon test in line with \cite{Flatscherdiss}.}
    \label{tab:coupon}
\end{table}

\begin{table}[h]
    \centering
    \begin{tabular}{|c|c|c|c|c|c|} \hline 
         $E_{11}$ (MPa)&  $E_{22}$ (MPa)& $\nu_{12}$&  $\nu_{21}$& $G_{12}$ (MPa)\\ \hline 
         146000&  9000& 0.02096&  0.34& 4270\\ \hline
    \end{tabular}
    \caption{Elastic properties of the composites.}
    \label{couponelastic}
\end{table}

\begin{table}[h]
    \centering
    \begin{tabular}{|c|c|c|c|c|} \hline 
         $R_{11}^T$ (MPa)&  $R_{22}^T$ (MPa)&  $R_{11}^C$ (MPa)&  $R_{22}^C$ (MPa)& $R_{12}$ (MPa)\\ \hline 
         2100&  82&  1407&  249& 110\\ \hline
    \end{tabular}
    \caption{Strength properties of the composites.}
    \label{tab:couponstrength}
\end{table}

\begin{table}[h]
    \centering
    \begin{tabular}{|c|c|c|c|c|c|} \hline 
        $G_{C,f}^T$ $\left( \dfrac{N}{mm} \right)$&  $G_{C,m}^T$ $\left( \dfrac{N}{mm} \right)$& $G_{C,f}^C$ $\left( \dfrac{N}{mm} \right)$&  $G_{C,m}^C$ $\left( \dfrac{N}{mm} \right)$& $\zeta_f $ & $\zeta_m$   \\ \hline 
         89.8&  0.2& 78.3&  0.8& 1&  1\\ \hline
    \end{tabular}
    \caption{Multi-phase-field properties.}
    \label{tab:couponPF}
\end{table}

This section discusses the results of the coupon tests in uni-axial compression and tension loading. The test considers the uni-axial loading of five angled-ply laminates as reported in \cite{Flatscherdiss}. The tensile and compressive stress states are compared with the experiments in line with \cite{Flatscherdiss}. Specifically, the layup sequence of $[(\theta/ -\theta)_8]_s$ with $\theta = 45^\circ,\ 60^\circ,\ 75^\circ$ are considered. In this section, the compressive tests are denoted with $C$, while the tension tests are denoted as $T$. For example, $ C-45$ presents the results of the uni-axial compression of the laminate with layup sequence  $[(45^\circ/-45^\circ)_8]_s$. The laminates are made of unidirectional Cycom\textregistered977-2-35/40-12KHTS-134-300 plies, a carbon-fiber epoxy laminate. The material properties for the simulations are taken from \cite{Flatscherdiss} and are presented in Tab. \ref{couponelastic} and \ref{tab:couponstrength}. Additionally, different fracture toughness was used in the experiment for tension and compression \cite{Flatscherdiss}, consequently adopted as shown in Tab. \ref{tab:couponPF}. Furthermore, the fiber and inter-fiber phase-field characteristic length scales are chosen as $\ell_f = 2 \; \mathrm{mm}$ and $\ell_m = 1 \; \mathrm{mm}$ respectively.

\begin{figure}[h]
\begin{center}
\includegraphics[width=0.90\linewidth]{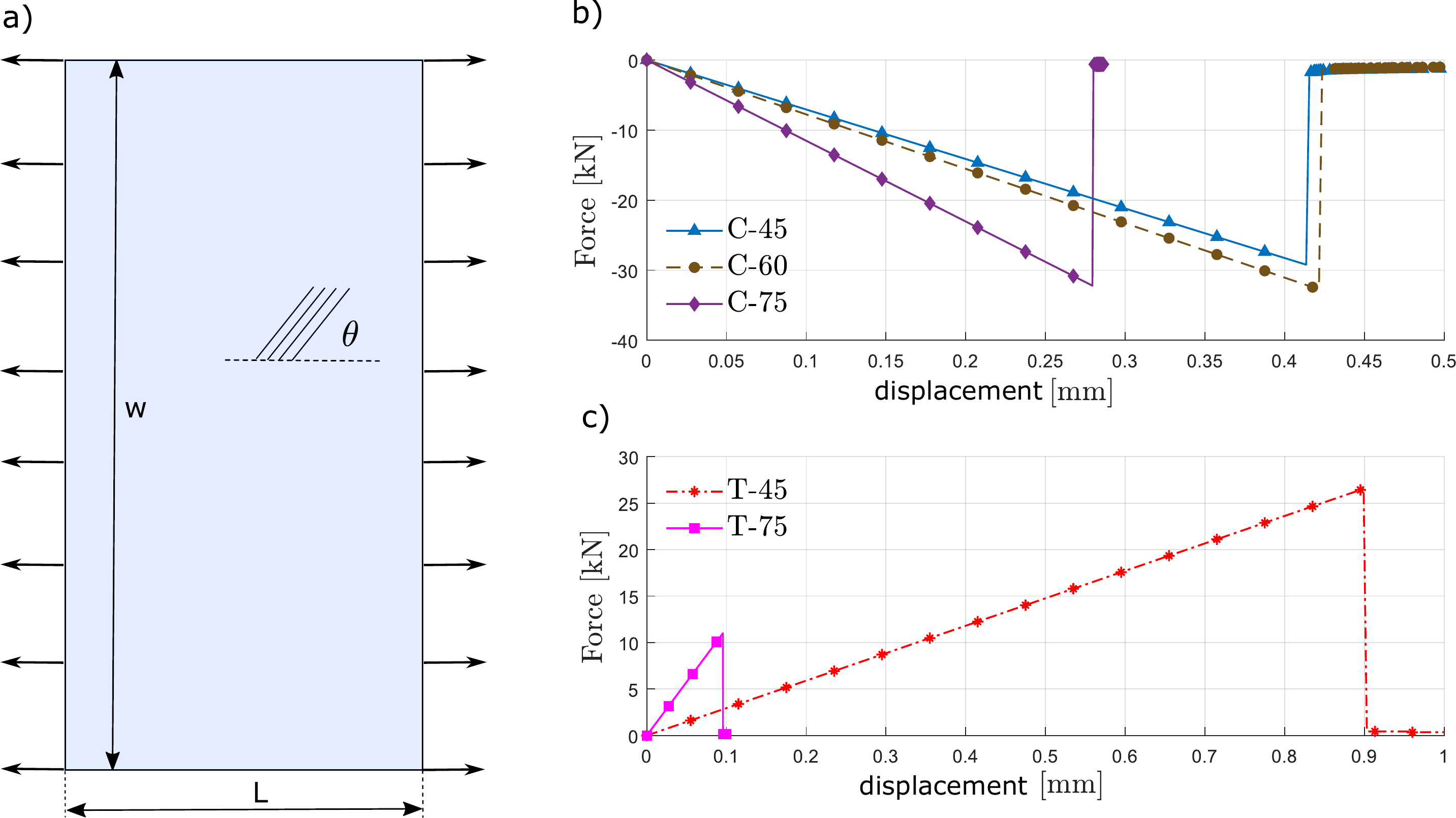}
\caption{a) Geometric description of the Coupon test along with the dimensions and boundary conditions (in tension), b) force vs. displacement curves corresponding to the compressive cases, and c) force vs. displacement curves corresponding to the tensile tests.}
\label{laminatecouponfd}
\end{center}
\end{figure}

Figure \ref{laminatecouponfd}a) presents the geometrical description of the bi-axial coupon tests under tension while the loads are reversed for the compression. Specifically, Table. \ref{tab:coupon} presents the layup sequence specific geometrical dimension with length $L$, and width $W$ in line with Fig.  \ref{laminatecouponfd}a). Each ply is considered $0.125 \ \mathrm{mm}$ thick, leading to the total thickness of laminates as $4$mm. 

\begin{figure}[h]
\begin{center}
\includegraphics[width=0.75\linewidth]{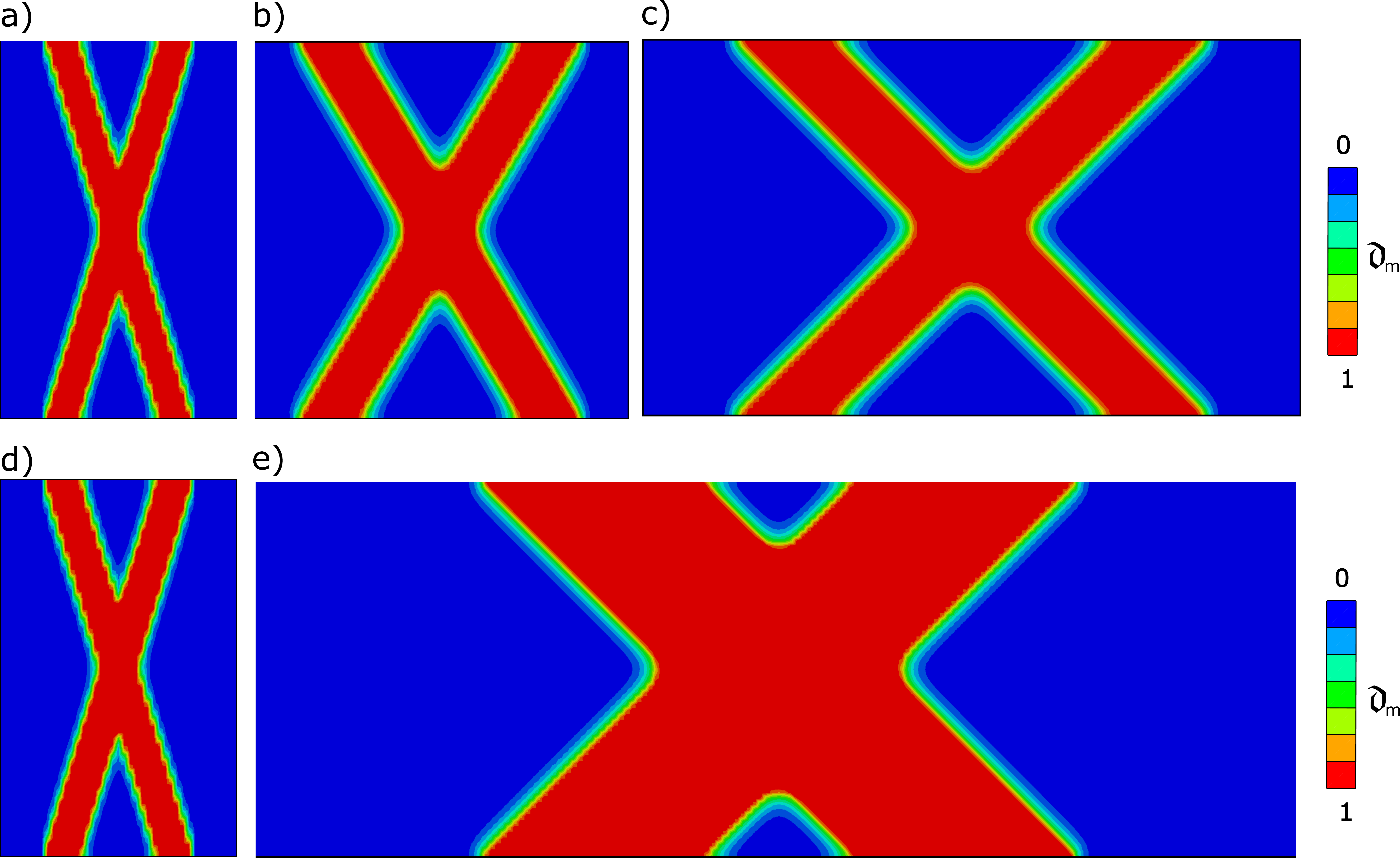}
\caption{Matrix crack pattern of the coupons for the compressive cases a) C-75, b) C-60, and c) C-45 as well as well as for the tensile tests d) T-75 and e) T-45.}
\label{laminatecoupon}
\end{center}
\end{figure}

In order to simulate the crack initiation and propagation, the laminate coupons are weakened in a small area near the centre of the coupons (of 4 elements) by reducing the fracture toughness of both fiber and inter-fiber by 10\% in line with \cite{N7, N4, N3}. Consequently, cracks are initiated in this region. 

Figure \ref{laminatecoupon}a), b) and c) presents the damage propagation in $ C-75$, $C-60 $, and $C-45$ respectively, while Fig.  \ref{laminatecoupon}d), and e) presents the damage propagation in $T-75$ and $ T-45$ respectively. Each case shows inter-fiber-dominated failure along the principle material orientation with no fiber damage. Furthermore, in all the cases, Puck matrix initiation is immediately followed by a symmetric unstable crack propagation. The resulting damage pattern for the $ T-45$ and $C-45$ are much broader than the other coupons , which could be attributed to  the effect of the structural tensor as detailed in \cite{Revisiting}. The force vs. displacement curve for the compression and tension cases are presented in Fig. \ref{laminatecouponfd} b) and c), respectively. All the force vs. displacement curves show a linear elastic behavior until the maximum load limit due to the absence of plasticity, followed by an unstable crack propagation indicated by a sudden drop in the reactions. Furthermore, the maximum load limit and the nominal cross section with width w, and thickness $t$ are used to compute the laminates stresses $ \bar{\sigma}_{xx}$ as in \cite{Flatscherdiss}:

\begin{equation}
    \bar{\sigma}_{xx} = \frac{F}{\text{w} t}.
\end{equation}

\begin{table}[h]
    \centering
    \begin{tabular}{|c|c|c|} \hline 
         Case&  Current Implementation (MPa) & Flatscher \cite{Flatscherdiss} (MPa) \\ \hline 
         C-45 &  $-228.2$ & $-203.1$\\
         C-60 &  $-255.7$ & $-198.4$\\
         C-75 &  $-251.6$ & $-229.3$\\
         \hline
    \end{tabular}
    \caption{Comparison of the nominal stress $\bar{\sigma}_{xx}$ determined by the current implementation and the implementation in \cite{Flatscherdiss}.}
    \label{flatschercomp}
\end{table}

Table  \ref{flatschercomp} compares the laminate stresses in the present model and the experimental investigation reported in \cite{Flatscherdiss}. The comparison in Tab \ref{flatschercomp} shows that the present implementation consistently overestimates the prediction in   \cite{Flatscherdiss} but generally proves to be in good agreement, especially in the absence of plasticity in the present model. Since the Puck failure initiation is immediately followed by the crack propagation in all the coupon tests,  the maximum load response is independent of the length scales $\ell_f $, and  $\ell_m $.  


\subsection{Open Hole Tension}\label{OHT}

The open hole tension has been addressed numerically using the phase-field method in \cite{ZHANG,trisha,Mrunmayee,HALLETT2009613,BLEYER2018213,FELGER201714} in a single ply setting. The open hole tension has been addressed numerically in the context of CDM as in\cite{KAUL2021100097,FLATSCHER20121090}. This section adopts the open hole tension results reported by Flatscher et al. \cite{FLATSCHER20121090,Flatscherdiss}. 

\begin{figure}[h]
\begin{center}
\includegraphics[width=0.750\linewidth]{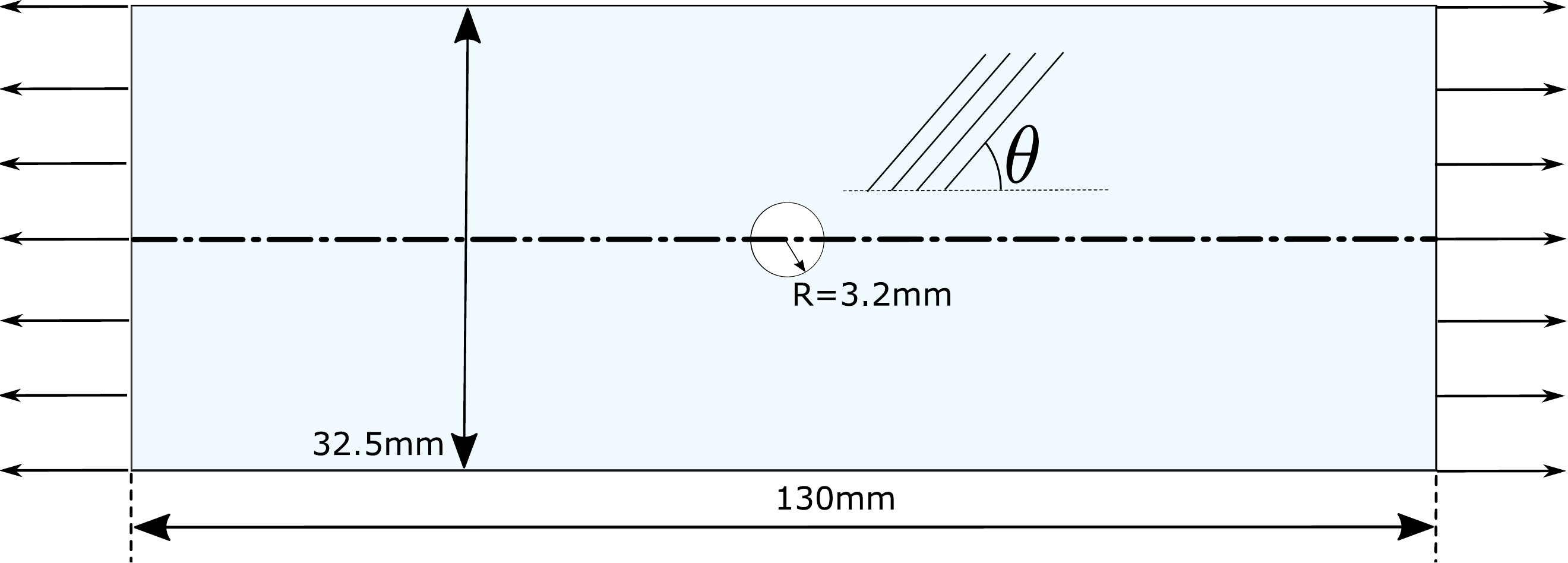}
\caption{Geometric description of open hole tension test along with the Boundary conditions, fiber orientation. The indicated symmetry plane (dash-dotted line) is used only for the $[0^\circ_4/90^\circ_4]_s$ layup.}
\label{fig:general_openhole_bc}
\end{center}
\end{figure}

The geometrical description of the open hole tension is presented in Fig. \ref{fig:general_openhole_bc}. Specifically, the two laminate sequences of $[0^\circ_4/90^\circ_4]_s$ and $[45^\circ_4/-45^\circ_4]_s$ are investigated. The material property is the same as in the section \ref{couponlaminates}  except the phase-field characteristic length scales, whose specific details will be presented in the sequel. 

\subsubsection{Open Hole Tension- {$[0^\circ_4/90^\circ_4]_s$}}

\begin{figure}[h]
    \centering
    \includegraphics[width=1\linewidth]{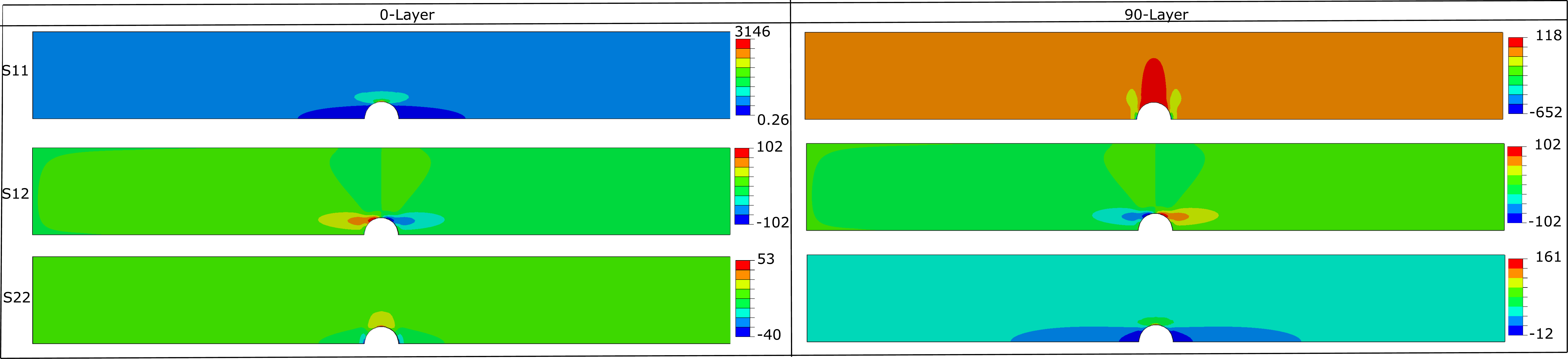}
    \caption{Stress distribution at each layer of the cross-ply laminates.  }
    \label{Stress_0_90}
\end{figure}

\begin{figure}[h]
\begin{center}
\includegraphics[width=0.50\linewidth]{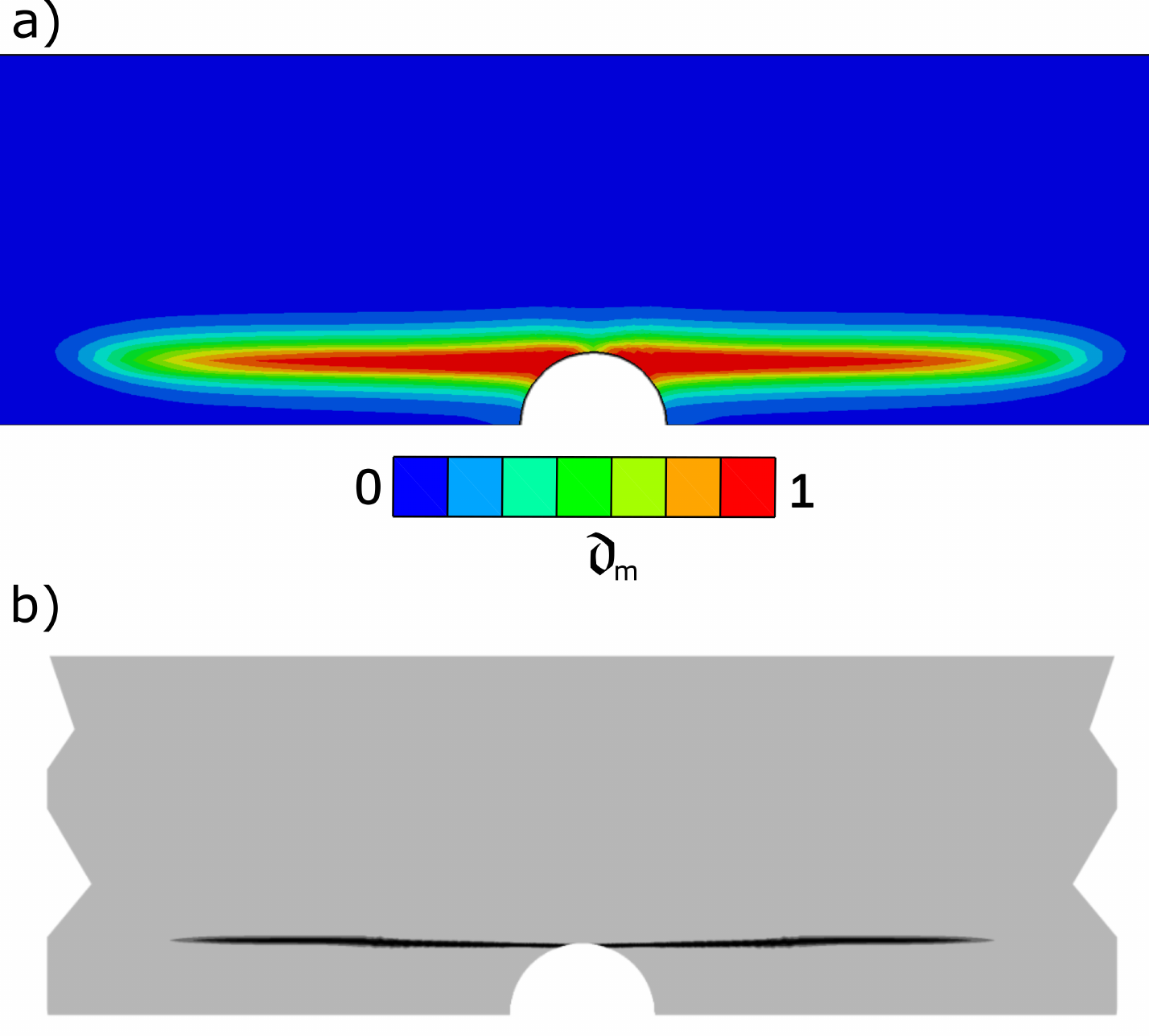}
\caption{Resulting inter-fiber damage pattern in a) the current implementation, and b) representative experimental results as in \cite{Flatscherdiss}.}
\label{fig:oht_090s_dmg}
\end{center}
\end{figure}

\begin{figure}[h]
\begin{center}
\includegraphics[width=0.95\linewidth]{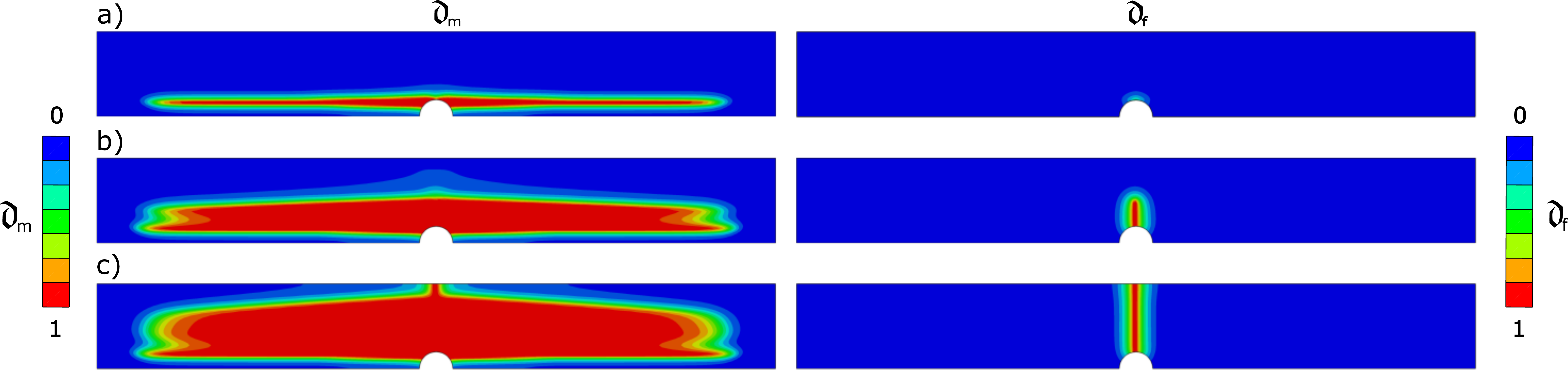}
\caption{Inter-fiber (left) and fiber (right) phase-field for a displacement of the right boundary of a) $0.7056\ \mathrm{mm}$, b) $0.7670\ \mathrm{mm}$, and c) $1\ \mathrm{mm}$.}
\label{fig:090_dmg_laterstages}
\end{center}
\end{figure}

Experimental work and the previous numerical methods \cite{FLATSCHER20121090,brogermaster,KAUL2021100097} regarding the open hole tension of $[0^\circ_4/90^\circ_4]_s$ suggest stiff load response since the fibers are oriented in the loading direction. Furthermore, the literature suggests that symmetric inter-fiber damage is propagated in the loading direction, followed by fiber damage in the direction perpendicular to the load. Consequently, only half of the specimen is modeled with symmetric boundary conditions. The domain is discretized by 12907 quadrilateral 4-node plane stress elements per ply block. Figure \ref{fig:general_openhole_bc} shows the symmetry plane used in this case. The vertical symmetry is not exploited due to the potential vertical crack. The fiber and inter-fiber's phase-field characteristic length scales are considered as $\ell_f = 2 \ \mathrm{mm}$ and $\ell_m = 1 \ \mathrm{mm}$ respectively. The dimensionless driving parameter for damage propagation are set to $\zeta_f = 30$ for fiber damage and $\zeta_m = 0.01$ for matrix damage.

Figure \ref{fig:oht_090s_dmg} compares the inter-fiber dominated failure in the present model and the experimental investigation. Furthermore, Figure \ref{Stress_0_90} presents the stress distribution in the individual layer. Initially, the stress concentrates near the hole, leading to inter-fiber-dominated damage around the vicinity of the hole. However, the inter-fiber-dominated damage starts to localize near the hole and then propagates parallel to the loading direction. The location of the inter-fiber dominated crack in Fig. \ref{fig:oht_090s_dmg}a) shows an excellent correlation with the experiments as in  Fig. \ref{fig:oht_090s_dmg}b). This predictive damage pattern can be interpreted as Longitudinal splitting in line with the experimental testing, see \cite{Yang2005,FLATSCHER20121090}· 

\begin{figure}[h]
\begin{center}
\includegraphics[width=0.99\linewidth]{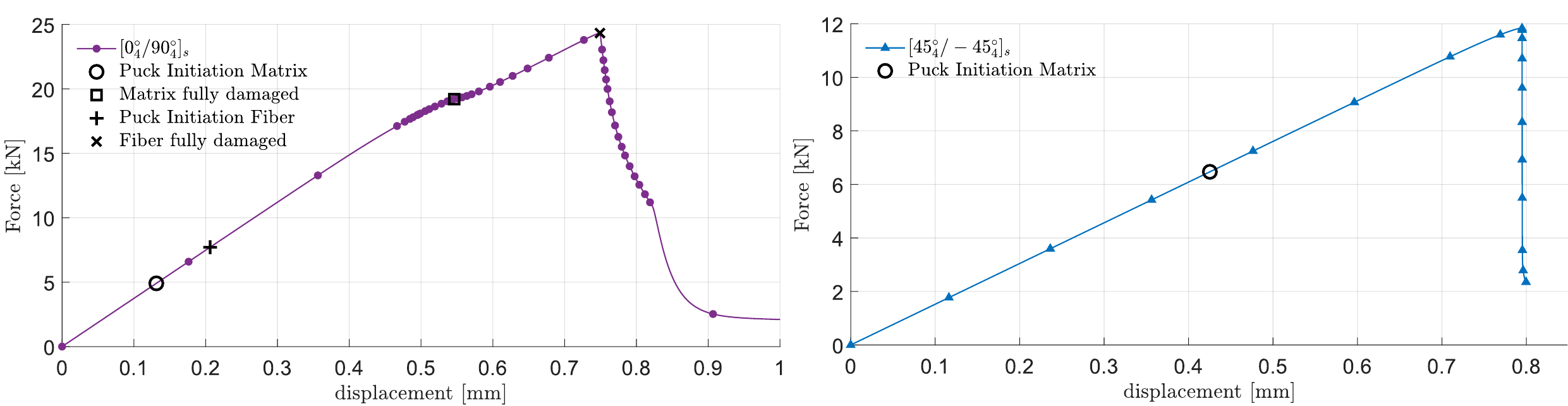}
\caption{Force displacement curve for an open-hole tension test with a layup sequence of a) $[0^\circ_4/90^\circ_4]_s$, and b) $[45^\circ_4/-45^\circ_4]_s$ along with the Puck damage initiation.}
\label{fig:090fdcurve}
\end{center}
\end{figure}

Figure \ref{fig:090fdcurve} presents the force vs displacement plot, Puck fiber, matrix initiation, and damage. In general, the force-displacement behavior is linear at first. A slight drop in the reaction force is observed due to strain location, which leads to an inter-fiber-dominated failure. The cracks propagate in the loading direction as shown in Fig.  \ref{fig:090_dmg_laterstages}a)left. At this stage, the Puck fiber damage is initiated, with no damage localization as shown in Fig.  \ref{fig:090_dmg_laterstages}a) right. The resulting damage pattern is compared to the numerical results reported in \cite{Flatscherdiss,KAUL2021100097}. Once the fiber damage is localized, the fiber damage propagates in the direction perpendicular to the loading direction as shown in Fig. \ref{fig:090_dmg_laterstages}b)right. Consequently, the inter-fiber dominated failure grows around the broken fibers as shown in Fig.  \ref{fig:090_dmg_laterstages}b) left. Furthermore, the full fiber damage leads to a complete structural failure as shown in  \ref{fig:090_dmg_laterstages}c). The resulting stable fiber damage can also be seen in Fig. \ref{fig:090fdcurve} by a steady drop in the force reactions.

The maximum load-carrying capacity of the present implementation is compared against the experimental and numerical investigation presented in \cite{FLATSCHER20121090} and \cite{brogermaster}. Table  \ref{tab:peakload090} presents the quantitative comparison of the peak load in the $[0^\circ_4/90^\circ_4]_s$ laminates. The load at failure in the experiment is estimated as $25.2 \; \mathrm{kN}$. The numerical results overestimate this peak load, while the current implementation matches the experiments very well, as shown in Tab \ref{tab:peakload090}. It is observed that change in the characteristic length of the phase-field $\ell_f $ and $ \ell_m$ does not influence the peak load, while the dimensionless driving parameters have a significant effect.

\begin{table}[h]
    \centering
    \begin{tabular}{|c|c|} \hline 
         Models &  Peak load (kN)\\ \hline 
         Experimental result \cite{FLATSCHER20121090} &  $25.2$\\
         Flatscher \cite{FLATSCHER20121090} &  $38.4$\\
         Broger \cite{brogermaster} &  $39.2$\\
         Current implementation &  $24.3$\\
         \hline
    \end{tabular}
    \caption{Comparison of the peak loads determined by the current implementation, the implementations in \cite{FLATSCHER20121090, brogermaster} and the experimental result \cite{FLATSCHER20121090}.}
    \label{tab:peakload090}
\end{table}






\subsubsection{Open Hole Tension- $[45^\circ_4/-45^\circ_4]_s$}

The open hole tension in the laminate $[45^\circ_4/-45^\circ_4]_s$ exhibit crack shielding \cite{KAUL2021100097}, leading to an asymmetric crack propagation. Consequently, the entire specimen is modeled using 56068 quadrilateral 4-node plane stress elements per ply with refinement around the open hole. The problem's geometrical description and boundary conditions are shown in Fig. \ref{fig:general_openhole_bc}. It has also been noticed that the damage localization occurs in a very small region. Consequently, a smaller characteristic length scale for the phase-field with $\ell_f = 0.2 \ \mathrm{mm}$ and $\ell_m = 0.1 \ \mathrm{mm}$ are chosen to resolve the gradient of the phase-field. Furthermore, the dimensionless driving parameter is considered to be $\zeta_f = 1$ and $\zeta_m = 0.005$. It is also important to note that the dimensionless parameters corresponding to the fiber damage are irrelevant since no fiber damage is observed as in \cite{FLATSCHER20121090,brogermaster,KAUL2021100097}.  

\begin{figure}[h]
\begin{center}
\includegraphics[width=0.80\linewidth]{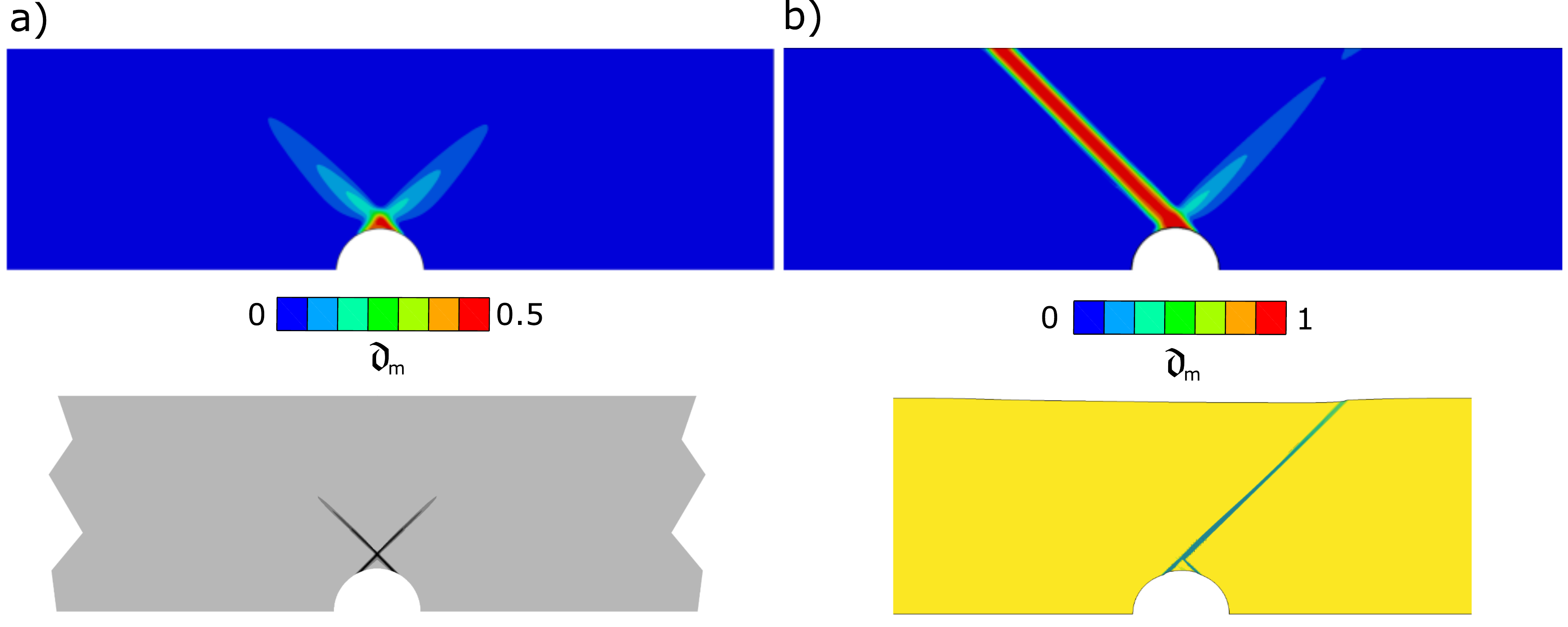}
\caption{ Comparison of a) initial damage to the result in \cite{Flatscherdiss} and b) the complete damage pattern to the result in \cite{KAUL2021100097}.}
\label{fig:45-45sdmg}
\end{center}
\end{figure}

\begin{figure}[h]
    \centering
    \includegraphics[width=0.7\linewidth]{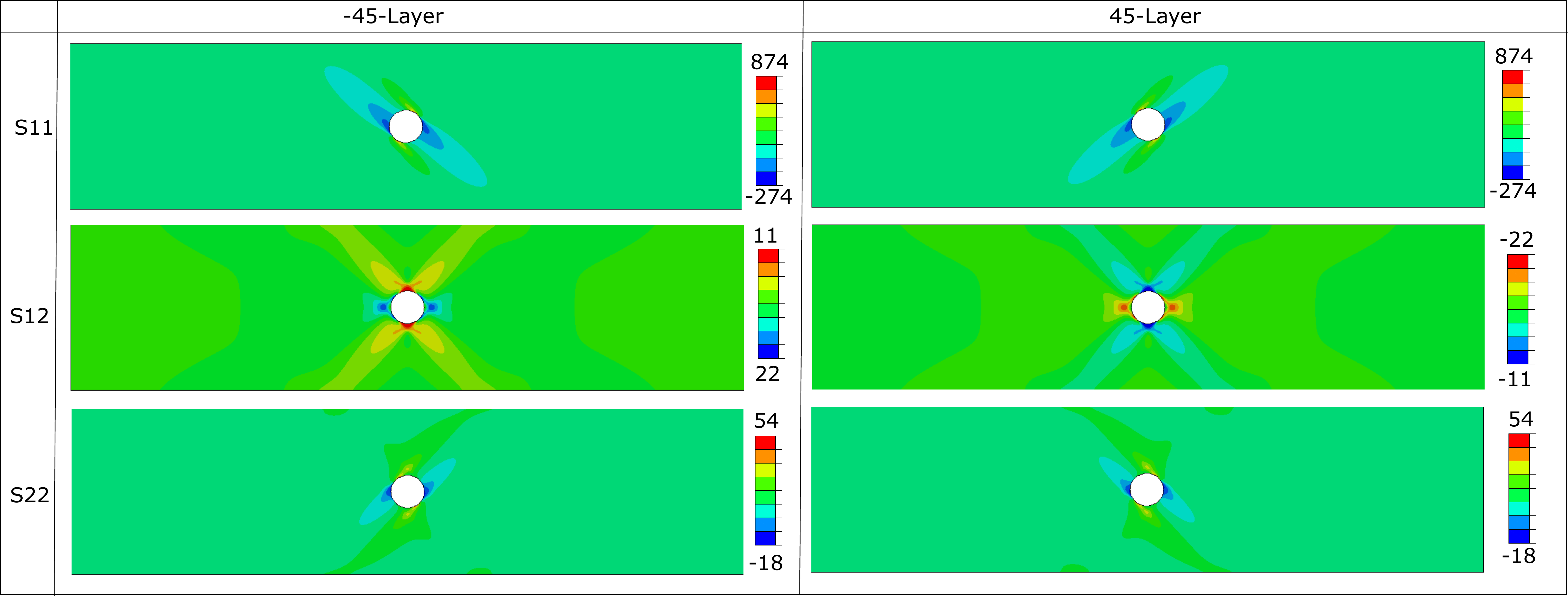}
    \caption{Stress distribution at each layer for the laminate of layup sequence $[45^\circ_4/-45^\circ_4]_s$. }
    \label{Stress_OHT_45}
\end{figure}

Figure \ref{fig:45-45sdmg} shows the different stages of damage accumulation demonstrated by the proposed implementation with comparisons to the results of \cite{Flatscherdiss, KAUL2021100097}. Meanwhile, Fig. \ref{Stress_OHT_45} presents the stress in each layer. Initially, the damage starts around the hole due to stress concentration. The damage then starts to accumulate, leading to localization as shown in Fig. \ref{fig:45-45sdmg}a). One of the branches starts to progress while the other does not due to crack shielding as shown in Fig. \ref{fig:45-45sdmg}b)  which is also evident in the literature, see \cite{KAUL2021100097} for more details.


The present model shows an excellent correlation with the experimental work regarding predicting the qualitative results of crack propagation. The force vs. displacement for the open hole tension with the laminate sequence of $[45^\circ_4/-45^\circ_4]_s$ is presented in Fig. \ref{fig:090fdcurve}b) along with the Puck inter-fiber failure initiation. The crack propagation mostly happens parallel to the fiber direction with no subsequence damage to the fiber. Consequently, the reactions are linear until the maximum failure load is reached, followed by an unstable crack propagation (crack shielding) leading to a drop in the forced displacement. Previous experiments and numerical results talk about pronounced non-linearity due to plasticity. As plasticity is not taken into account in this model, these non-linearities are not reflected in Fig. \ref{fig:090fdcurve}b). Table \ref{tab:peakload45-45} compares the predicted peak load of the present model with the experimental and numerical observation from the literature. It is seen that the present model predicts the peak load very well.

\begin{table}[h]
    \centering
    \begin{tabular}{|c|c|} \hline 
         Models &  Peak load (kN)\\ \hline 
         Experimental result \cite{FLATSCHER20121090} &  $14.1$\\
         Flatscher \cite{FLATSCHER20121090} &  $10.7$\\
         Broger \cite{brogermaster} &  $9.0$\\
         Current implementation &  $11.8$\\
         \hline
    \end{tabular}
    \caption{Comparison of the peak loads determined by the current implementation, the implementations in \cite{FLATSCHER20121090,brogermaster} and the experimental result \cite{FLATSCHER20121090}.}
    \label{tab:peakload45-45}
\end{table}

\subsection{Compact Tension} \label{compacttens_lam}

\begin{figure}[h]
\begin{center}
\includegraphics[width=0.50\linewidth]{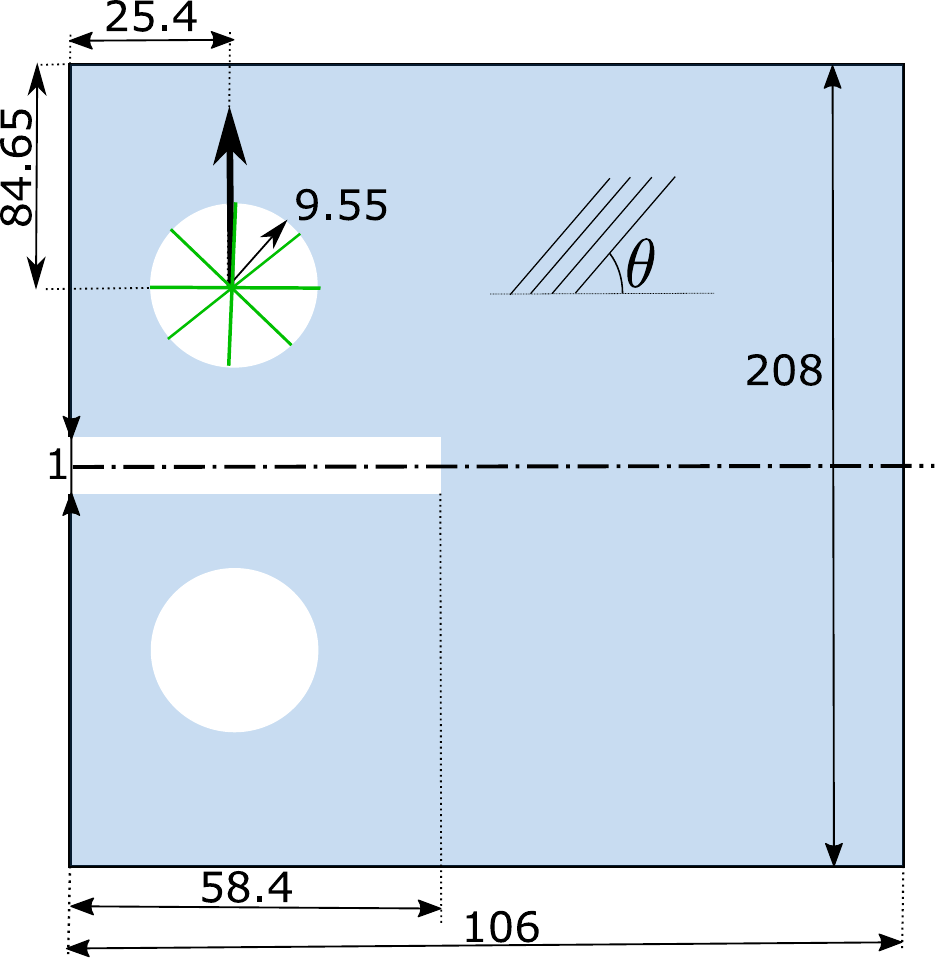}
\caption{Geometrical description of contact tension tests along with the  boundary conditions. The symmetry plane (dash-dotted line) is used for this simulation. The green lines indicate a rigid connection of the surface to the central node which experiences the displacement. All lengths are considered to be in mm.}
\label{fig:ctlam_bcfd}
\end{center}
\end{figure}

This section corresponds to the compact tension results of the extended compact tension proposed in  Li et al. \cite{LI20091891}. Li et al. \cite{LI20091891} investigate the influence of layup sequence on the fracture behavior of the laminated composites using blocked ply and dispersed ply sequences. The material consists of an IM7/8552 carbon-epoxy pre-preg system with a nominal ply thickness of $0.125$ mm. While both types of layup sequences consist of the same number of plies in each laminate,  for the dispersed plies, each ply had a different fiber orientation to the next ply. In contrast, the blocked ply layup sequences introduced thicker sheets of one fiber orientation by stacking more plies with the same fiber orientation above each other, leading to a stark contrast in their fracture behavior. See \cite{LI20091891} for more details. The dispersed ply specimens showed clear fiber failure and only moderate inter-fiber-dominated damage, with fiber failure occurring in multiple unstable events. However, the blocked ply specimens were more prone to delamination and matrix damage, leading to a pronounced non-linear response in their force-displacement behavior. Furthermore, the quasi-isotropic specimens exhibit more complex fracture behavior than the cross-ply samples. The difference in the layup sequence cannot be captured with the mesh overlay method. Therefore, only blocked cross-ply results are compared to the proposed model.\\

\begin{figure}[]
\begin{center}
\includegraphics[width=0.70\linewidth]{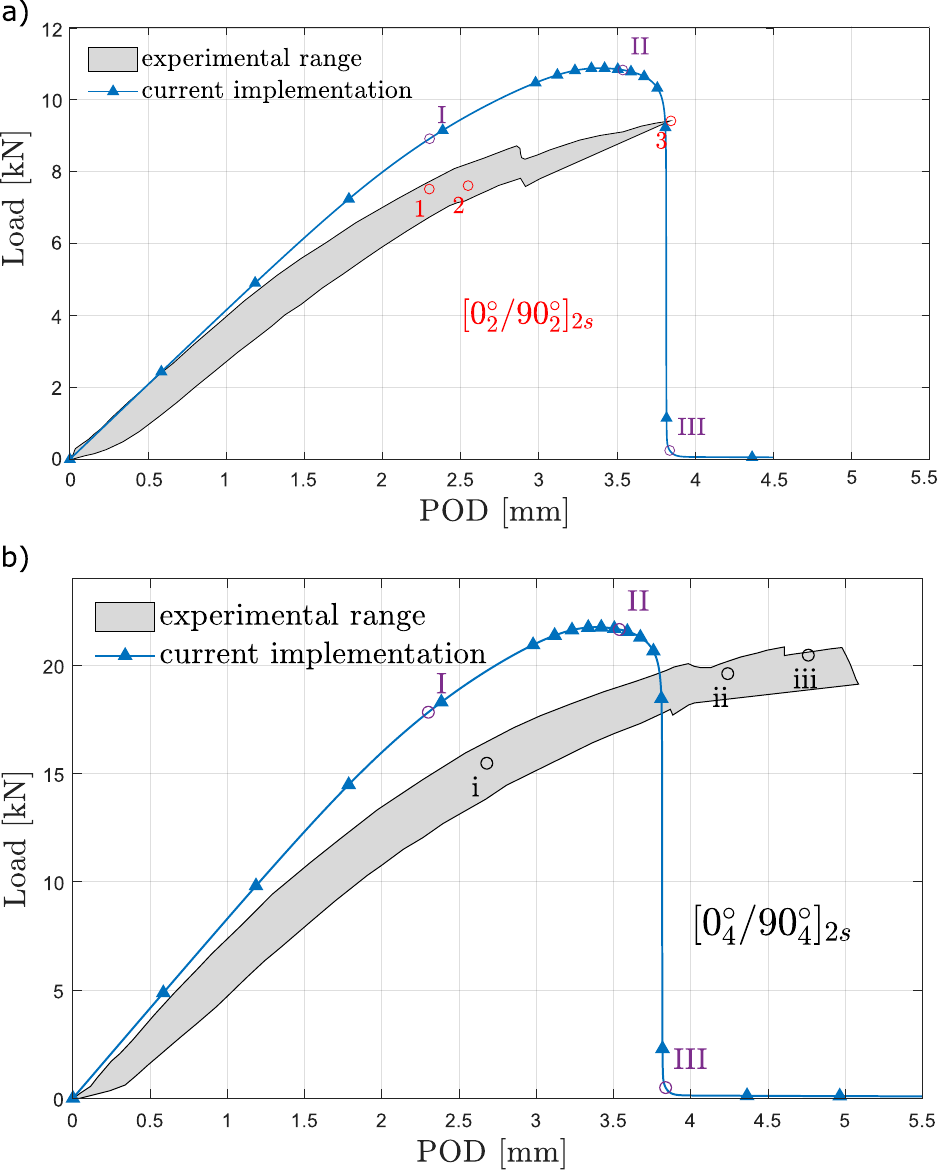}
\caption{a) Force vs. Pin Opening Displacement (POD)  curves corresponding to the $[0^\circ_2/90^\circ_2]_{2s}$,  and b) force pin opening displacement curves corresponding to the $[0^\circ_2/90^\circ_2]_{2s}$ . The markings with numerals indicate the positions at which the images in Fig. \ref{fig:ctlam_pf} were taken.}
\label{fig:ctlam_fd_only}
\end{center}
\end{figure}

\begin{figure}[]
\begin{center}
\includegraphics[width=0.90\linewidth]{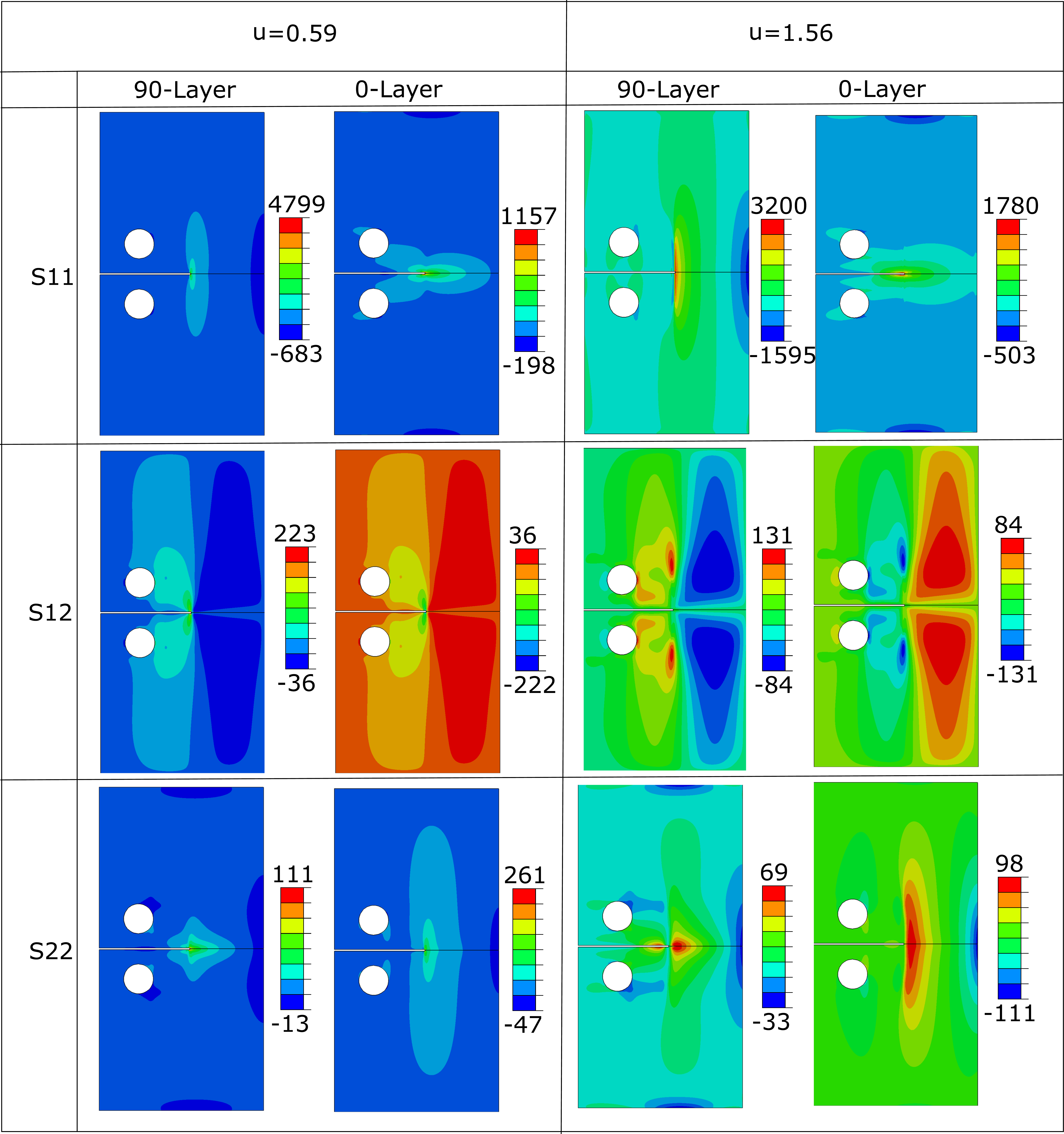}
\caption{Stress distribution at each layer of $[0^\circ_2/90^\circ_2]_{2s}$ corresponding to the displacement at position I ($d=0.59$) and II ($d=1.55$) respectively.}
\label{fig:stress_CT}
\end{center}
\end{figure}

Figure \ref{fig:ctlam_bcfd} a) presents the geometrical description of the body under consideration and the boundary conditions. In order to mimic the experiments as closely as possible, the displacement is applied so that the hole can rotate around a central node without deforming. The domain is discretized by $10620$ quadrilateral 4-node plane stress elements per ply.

\begin{figure}[]
\begin{center}
\includegraphics[width=0.90\linewidth]{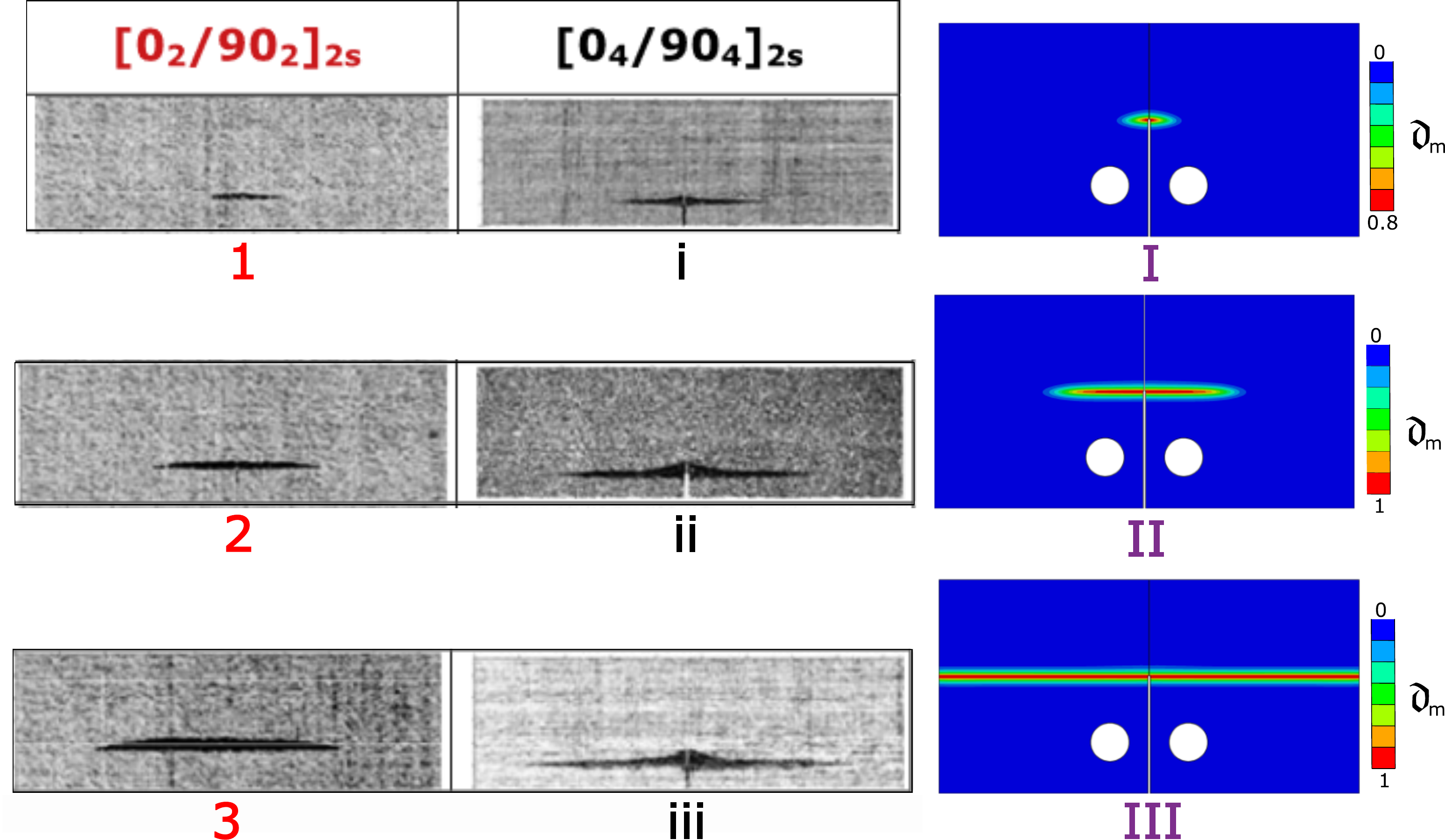}
\caption{Comparison of the crack propagation of the numerical results (purple roman numerals) to the experimental results (red arabic and small black roman numerals). Experimental results are reproduced from \cite{LI20091891} with permission from Elsevier.}
\label{fig:ctlam_pf}
\end{center}
\end{figure}

Quantitatively, the force versus pin opening displacement (POD) curve is compared to the measured results from \cite{LI20091891} in Fig.\ref{fig:ctlam_fd_only}a) for the  $[0^\circ_2/90^\circ_2]_{2s}$ and Fig.\ref{fig:ctlam_fd_only}b) for the $[0^\circ_4/90^\circ_4]_{2s}$ layups. The comparisons are based on a single numerical example of $[0^\circ/90^\circ]$. For the $[0^\circ_4/90^\circ_4]_{2s}$ case, the reaction forces were scaled accordingly to incorporate the increased thickness. Additionally, Fig. \ref{fig:ctlam_fd_only} show the points in the load-displacement curves, where the pictures in Fig. \ref{fig:ctlam_pf} were taken to facilitate a better qualitative comparison. The numerical result is always marked with purple Roman numerals, while the experimental results are marked with red and black Roman numerals.

Figures \ref{fig:ctlam_fd_only} a) and b) generally display a good agreement of the force-displacement behavior of the experiment, especially at low loading stages, since both the experiment and the numerical result show a linear relation between force and displacement with very similar slopes in both cases. Furthermore, the numerical result consistently overpredicts the reaction forces and shows a much longer loading period, where the curve remains linear. This discrepancy can be due to the heavy delamination between the plies that occur in the experiments. In spite of that, the present model generally matches the experimental results well.\\

The indications in Fig. \ref{fig:ctlam_fd_only} highlight the loading stages at which the damage presented in Fig. \ref{fig:ctlam_pf} is taken. The initial images 1, i, and I were taken at roughly the same displacement stage, while the second set of pictures differs. This is due to a seemingly unstable crack propagation event that happens in the experiment of the $[0^\circ_2/90^\circ_2]_{2s}$ case between pictures 1 and 2, leading to a larger crack compared to the numerical result at the same displacement. Therefore, picture II was taken closer to point ii at a latter loading stage. The last picture of the numerical result already shows complete failure, while the experiments do not reach complete failure as is the case for the $[0^\circ_2/90^\circ_2]_{2s}$ specimens or failure happens later as for the $[0^\circ_4/90^\circ_4]_{2s}$ samples. The latter fails by the pullout of the $0^\circ$-blocks instead of matrix cracking. This also explains the discrepancy in the final damage pattern predicted in Fig. \ref{fig:ctlam_pf}. Here, the energy required for the inter-fiber dominated crack to grow is dissipated by delamination and $ 0^\circ$-block pullout, leading to shorter matrix cracks than the numerical result. The stress at each layer corresponding to the damage I and II are presented in Fig. \ref{fig:stress_CT}.

In conclusion, the proposed model provides an excellent agreement with the experiments in terms of both qualitative and quantitative predictions.

\subsection{Double Notched Tension}\label{DNT}

This section corresponds to the double-edge notched tension results of various laminate sequences. The article \cite{Hallet_R52} presents the experimental results of the progressive damage and the effects of the layup sequence in the double-edge tension tests. The article uses specimens of various sizes to present the failure mode, strength, and subcritical damage development in the E-glass/913 glass-epoxy laminate. In this section, three laminate sequences with constant width $w=20$mm  are considered to evaluate the model's predictive capability in terms of fiber, inter-fiber failure interaction, sub-critical damage,  and crack propagation. 

\begin{table}[h]
    \centering
    \begin{tabular}{|c|c|c|c|c|c|} \hline 
         $E_{11}$ (MPa) &  $E_{22}$ (MPa) &  $E_{12}$ (MPa) & $\nu_{12}$ &  $\nu_{21}$ & $G_{12}$ (MPa)\\ \hline 
         45600 &  16200 &  6000 & 0.099 &  0.4& 6000\\ \hline
    \end{tabular}
    \caption{Elastic Properties of the glass-epoxy}
    \label{DN_elasstic}
\end{table}

\begin{table}[h]
    \centering
    \begin{tabular}{|c|c|c|c|c|l|l|} \hline 
         $R_{11}^T$ (MPa)&  $R_{22}^T$ (MPa)&  $R_{11}^C$ (MPa)&  $R_{22}^C$ (MPa)& $R_{12}$ (MPa) &$G_{C,f}^T$ $\left( \dfrac{N}{mm} \right)$ &$G_{C,m}^T$ $\left( \dfrac{N}{mm} \right)$
\\ \hline 
         1280&  73.3&  800&  145& 73 &64 &1.8\\ \hline
    \end{tabular}
    \caption{Strength properties of glass-epoxy}
    \label{DN_Strength}
\end{table}

\begin{figure}[h]
\begin{center}
\includegraphics[width=0.80\linewidth]{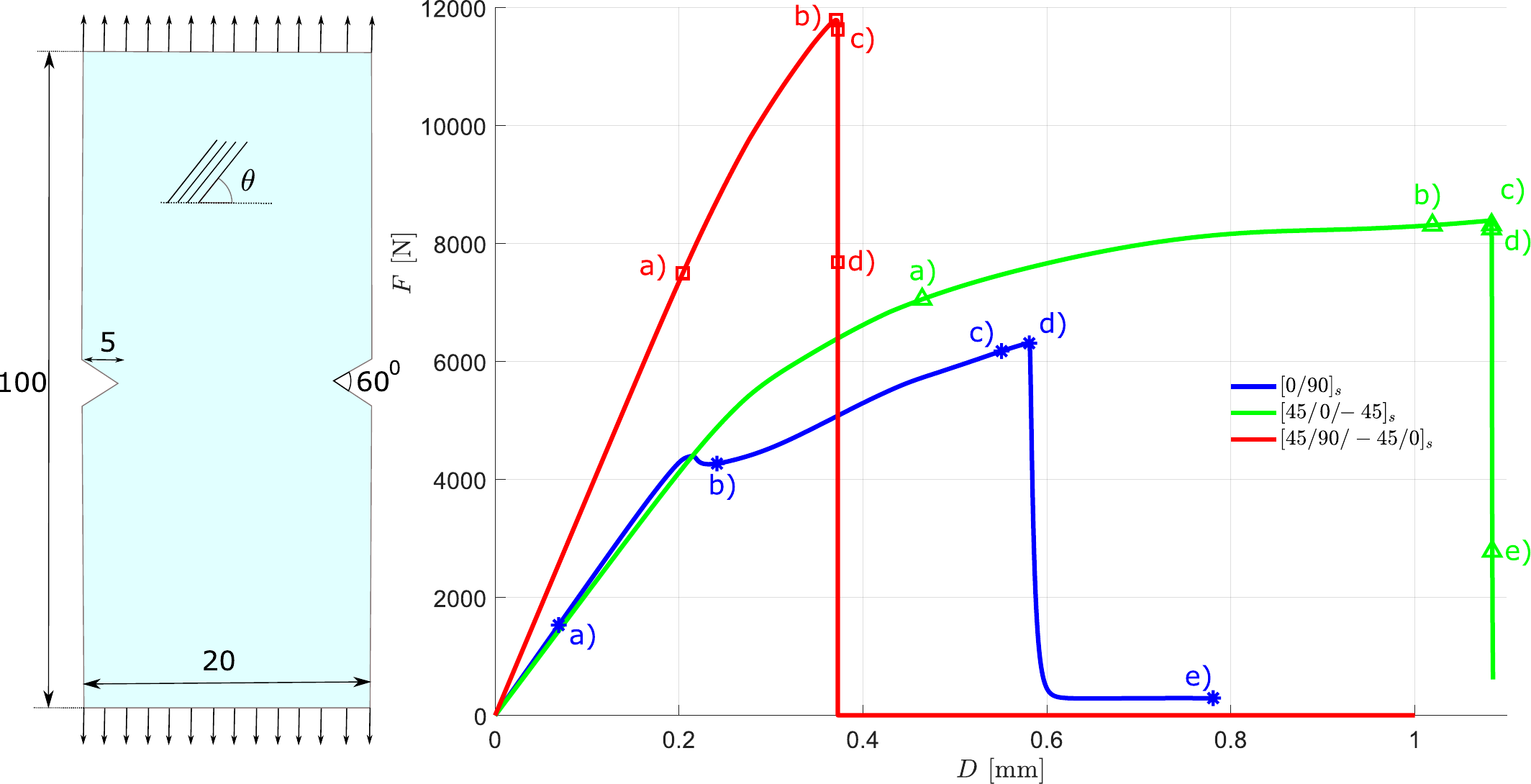}
\caption{a) Geometric description of the double-notched tension specimen along with the boundary conditions an fiber orientations, b) force vs. displacement curves for different laminate sequences.}
\label{DN_Diagram}
\end{center}
\end{figure}

The geometrical description of the body under consideration, along with the boundary conditions, is presented in Fig. \ref{DN_Diagram} a). The Figure  \ref{DN_Diagram} a) also presents the nomenclature of the fiber orientation $\theta$. The direction of loading is considered as $\theta=90^\circ$, while the direction perpendicular to the loading direction is considered as $\theta=0^\circ $. Three laminate sequences are considered in line with \cite{Hallet_R52}, including a) $[0/90]_s $, b) $[45/90/-45]_s$ (in this article, the sequence translates to $[45/0/-45]_s$), and c) $[45/90/-45/0]_s$ .

The material consists of glass-epoxy pre-preg plates whose approximate material properties are presented in the Table. \ref{DN_elasstic}-\ref{DN_Strength}.   The model is meshed with 23716 elements for each ply, with refinement near the notches in line with \cite{Hallet_numerics}. Each layup sequence is accompanied by a comparison with the experimental results \cite{Hallet_R52} at several instances in line with the experiments. Furthermore, a force-displacement curve for each layup sequence is presented, indicating the damage patterns at intervals as in \cite{Hallet_R52} is shown to highlight the model's ability to simulate sub-critical cracks, crack propagation in both fiber and inter-fiber separately. In all the laminated sequences presented in this section, the authors \cite{Hallet_R52} report extensive delamination, which is neglected in the numerical modeling.  

\subsubsection{Cross-Ply Laminate $[0/90]_s$}

\begin{figure}[h]
\begin{center}
\includegraphics[width=0.90\linewidth]{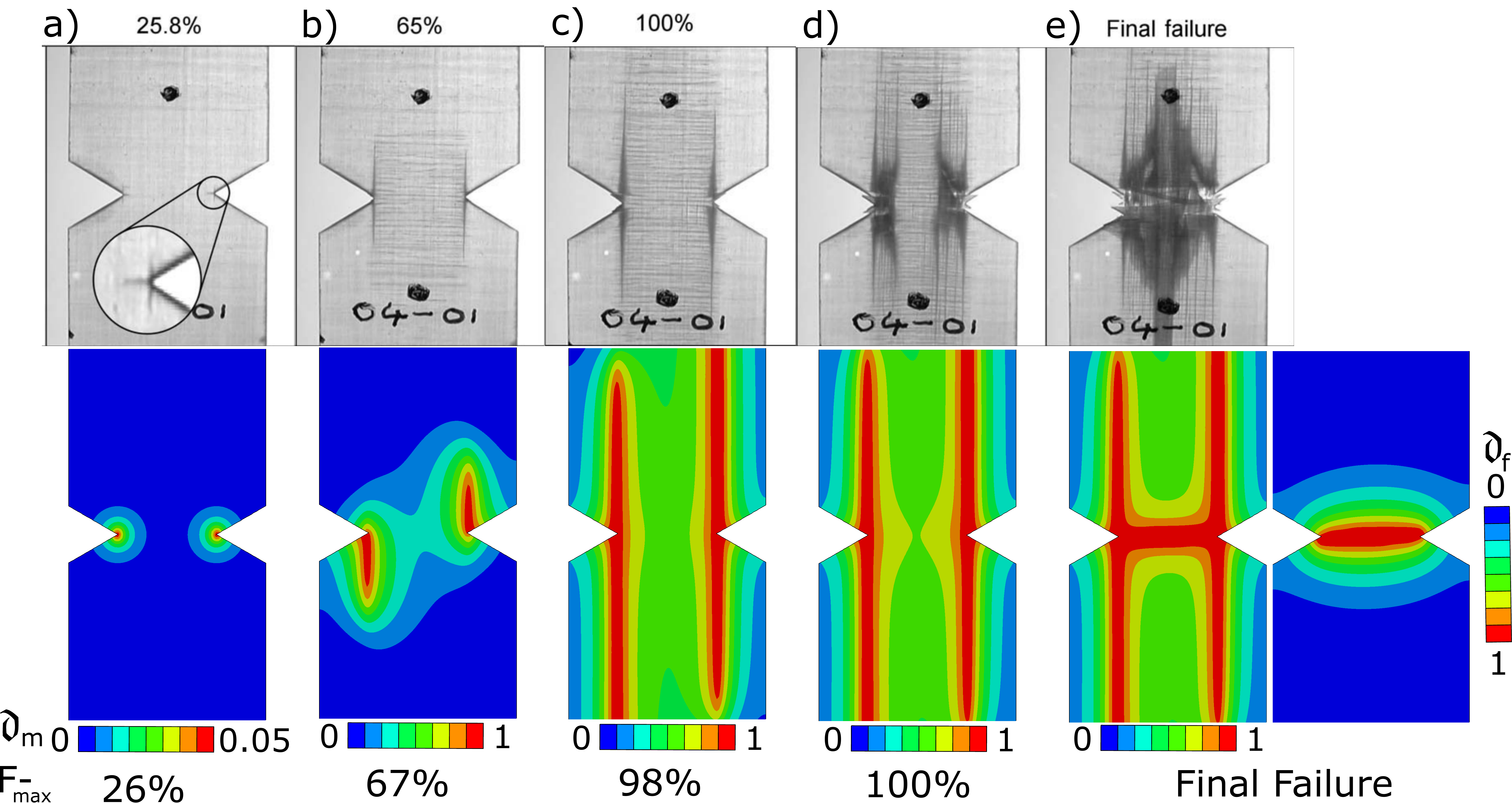}
\caption{Comparison of the crack propagation between the experiment and numerical results for the laminates sequence of $[0/90]_s$ as a function of maximum load. Experimental results are reproduced from \cite{Hallet_R52}}
\label{DN_crack_090}
\end{center}
\end{figure}

Figure \ref{DN_crack_090} compares numerical results and the experimental results of the cross-ply laminate in terms of crack propagation. Figure \ref{DN_Diagram}b) presents the force vs displacement curve for the cross-ply laminate. The percentage of the maximum reaction force is shown for each damage comparison, and the corresponding crack pattern is compared with the experimental results \cite{Hallet_R52}. The comparison of crack propagation at a) 26\% , b) 67\%, c) 98\%, d) 100\% of maximum load is presented in Figure \ref{DN_crack_090}a),b),c),d) respectively. The Final failure is presented in e).  

Initially, the split near the notch appears in the cross-ply laminate as shown in Fig. \ref{DN_crack_090}a). At this point, there is a partial inter-fiber failure, as shown in the numerical results. Furthermore, as the load increases, ply splitting occurs in the loading direction, while the single transverse cracks appear in the directions perpendicular to the loading direction. This can also be seen numerically in Fig. \ref{DN_crack_090}b). The split cracks propagate, while the region perpendicular to the loading starts to damage  $\pf_m \approx 0.5 $. Due to the inter-fiber dominated crack in the cross-ply, a slight drop in the force vs. displacement can be seen in Fig. \ref{DN_Diagram}b). As the load further increases, the splitting crack grows, leading to fiber damage near the notch tip as shown in Fig. \ref{DN_Diagram} c) and d). Experimental investigation \cite{Hallet_R52} suggests that as the fiber failure increases, it follows an irregular path across the specimen width and is accompanied by splitting and delamination. Omitting the delamination, Fig. \ref{DN_Diagram} d), e) presents the growth of a splitting crack followed by a fiber failure. As the fiber failure increases, the matrix around the fiber also damages, as shown in Fig. \ref{DN_Diagram} e).  

The fracture prediction based on individual ply can be made based on the stress concentrations in the individual ply. Fig. \ref{fig:ctlam_pf_stress} presents the stress distribution for the cross-ply laminate in each layer.

\subsubsection{Laminate $[45/0/-45]_s$}

\begin{figure}[h]
\begin{center}
\includegraphics[width=0.80\linewidth]{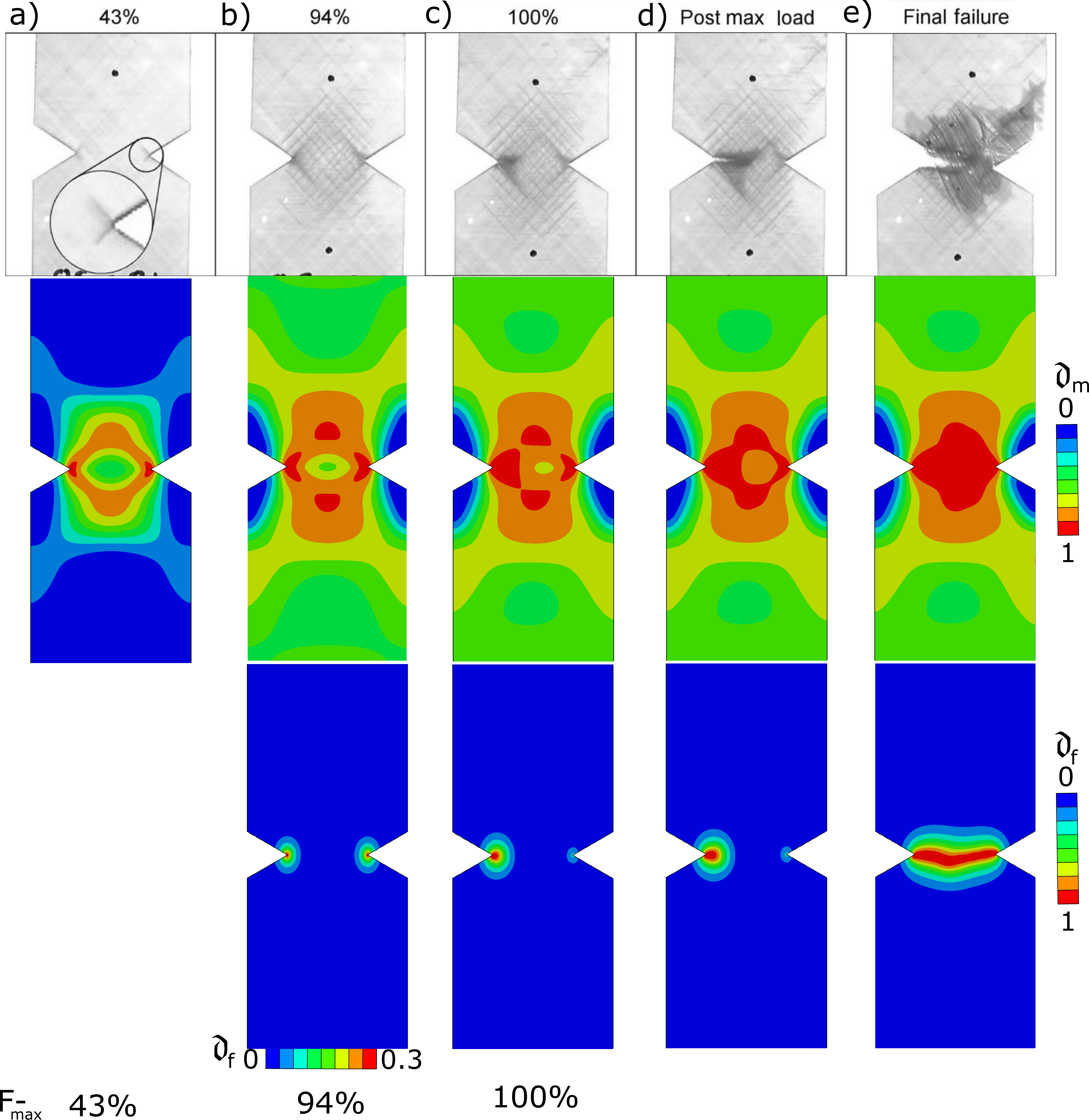}
\caption{Comparison between experimental and numerical results for the laminates sequence of $[45/0/-45]_s$ as a function of maximum load. The second row presents the damage in inter-fiber, while the third row presents the corresponding fiber-failure (if any). Experimental results are reproduced from \cite{Hallet_R52}}
\label{DN_crack4590_45}
\end{center}
\end{figure}

Figure \ref{DN_crack4590_45} compares the experimental and numerical investigation concerning crack propagation based on the percentage of maximum load. Fig. \ref{DN_Diagram}b) presents the corresponding force vs. displacement.

As the applied load increases, the inter-fiber dominant cracks at the $45^\circ $interface and $-45^\circ $ layers start to appear. The cracks are oriented in the fiber direction in the local setting, while the cracks appear to grow  $\pm 45^\circ $ in the global setting as shown in Fig. \ref{DN_crack4590_45}a). At this stage, the numerical results show damage in the laminates dominated by the $\pm 45^\circ $ inter-fiber. 

As the applied load increases, subcritical cracks start to nucleate to form a more pronounced crack pattern in an inter-fiber dominated failure mode as in Fig. \ref{DN_crack4590_45}b). The Numerical results also show that the combination of fiber and inter-fiber failure modes drives the damage after this stage. When the maximum reaction load is reached, the failure mode is entirely dominated by the fiber failure at the notch tip, as shown in Fig. \ref{DN_crack4590_45}c). Due to the stark rise in the fiber damage, the force vs. displacement shows a significant drop, leading to a complete failure as shown in Fig. \ref{DN_Diagram}b). The crack propagation and the final failure pattern are compared in Fig. \ref{DN_crack4590_45}d) and e), respectively. 

In conclusion, the results presented show an excellent correlation with the experimental results for the qualitative comparison. Specifically, the models predict an inter-fiber-dominated failure followed by a fiber-dominated failure. The model also captures the competition between the fiber and matrix-dominated failure mechanisms as shown in Fig. \ref{DN_crack4590_45} in line with the experiments \cite{Hallet_R52}.

\subsubsection{Quasi-Isotropic Laminate $[45/90/-45/0]_s$}

\begin{figure}[]
\begin{center}
\includegraphics[width=0.80\linewidth]{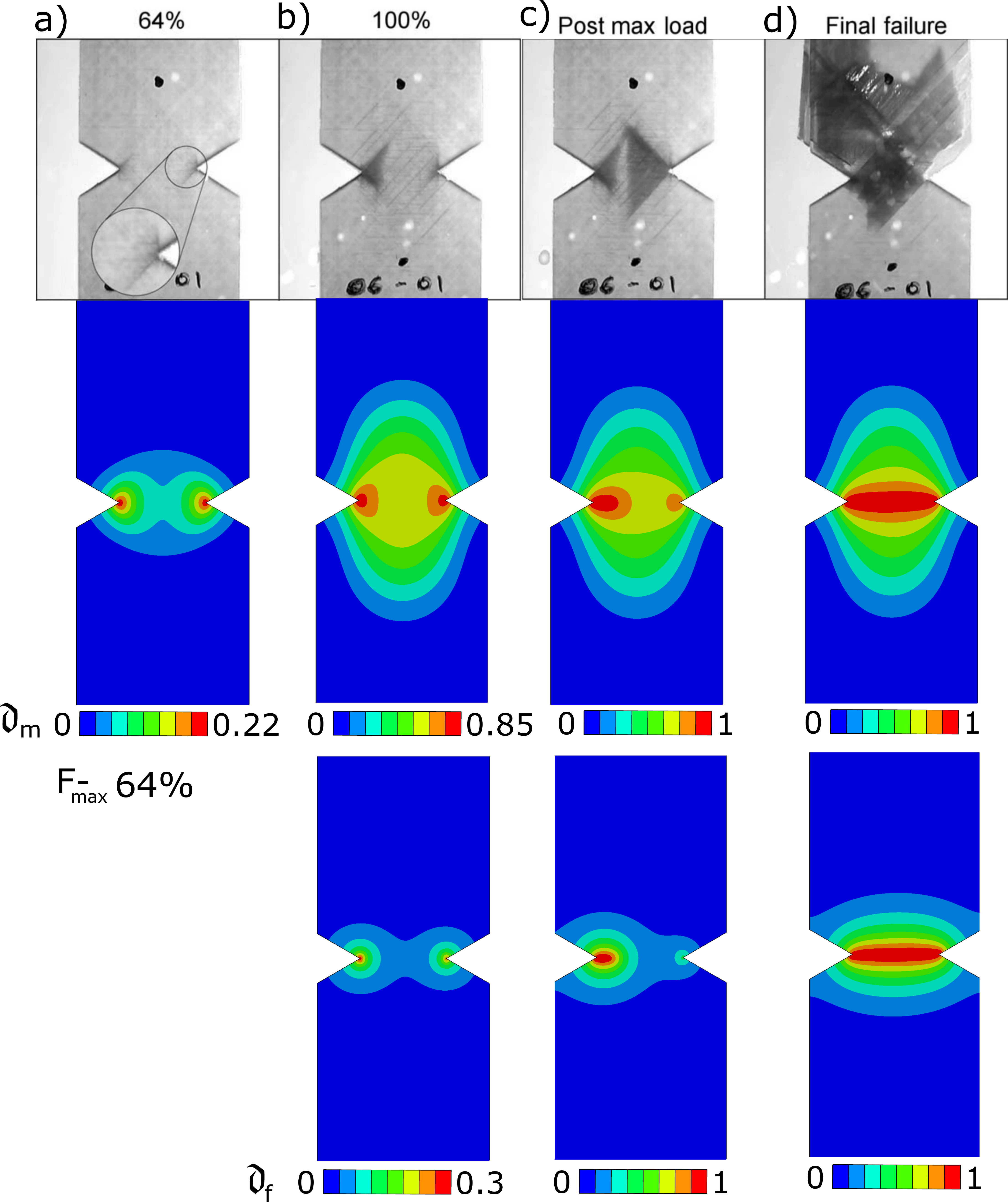}
\caption{Comparison between experimental and numerical results for the laminates sequence of $[45/90/-45/0]_s$ as a function of maximum load. The second row presents the damage in inter-fiber, while the third row presents the corresponding fiber-failure (if any). Experimental results are reproduced from \cite{Hallet_R52}}
\label{DN_crack4590_45_0}
\end{center}
\end{figure}

\begin{figure}[]
\begin{center}
\includegraphics[width=0.80\linewidth]{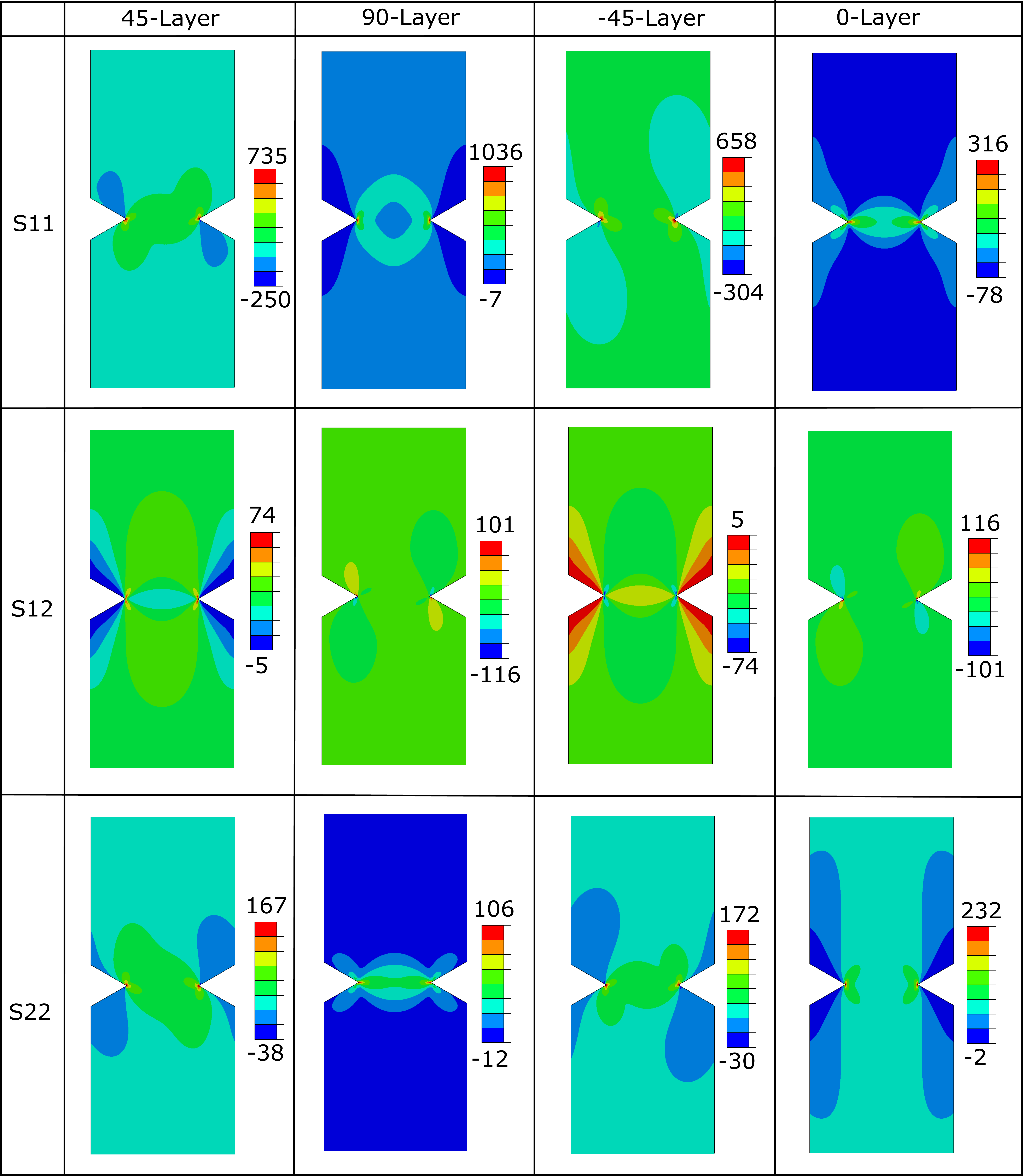}
\caption{Stress distribution in each layer of the laminate $[45/90/-45/0]_s$ corresponding to maximum load of $F_{max}=64\%$ corresponding to crack propagation in Fig. \ref{DN_crack4590_45_0}a). All the stresses are in MPa. }
\label{DN_stress4590_45_0}
\end{center}
\end{figure}

This subsection compares the quasi-isotropic laminates with the layup sequence of $[45/90/-45/0]_s$ with the experimental results. The results are similar to the laminate sequence of $[45/0/-45]_s $. The interlaminar dominant cracks run along the fiber direction of $\pm 45^\circ $ plies at first as shown in Fig. \ref{DN_crack4590_45_0}a). This subcritical crack further grows, and the competition between the inter-fiber and fiber-dominated failure can be seen in Fig. \ref{DN_crack4590_45_0}b). Furthermore, as the fiber-dominated failure increases, the crack evolution strongly depends on the fiber failure, with inter-fiber-dominated failure around the damaged fibers as shown in Fig. \ref{DN_crack4590_45_0}c) and Fig. \ref{DN_crack4590_45_0}d). Furthermore, the force vs. displacement curve for the quasi-isotropic layup is shown in Fig. \ref{DN_Diagram}b). The stress distribution in each layer at position a) is presented in Fig. \ref{DN_stress4590_45_0}.

\section{Conclusion}
\label{conlusions}

A multi-phase-field ply model capable of handling various laminate sequences is presented in this article. The model incorporates the Puck failure theory and uses two separate damage variables to separately model failure in fiber and inter-fiber. The laminate with different layup sequences is built using the mesh overlay technique. Comparison with the experiments showcases the model's ability to replicate qualitative and quantitative experimental results. Furthermore, the interplay between the fiber and inter-fiber damage is shown using the cross-ply and isotropic laminates. Additionally, the compressive failure of the laminates is addressed in the article. 

The present model can currently be used for the qualitative assessment of fractures in laminates. In contrast, the quantitative assessment can be made by calibrating the proposed dimensionless number coined as dimensionless driving parameters. Finally, this work brings together many benchmark examples of computational analysis of laminates. Additionally, the codes and the corresponding data are provided to bring transparency and allow independent researchers to build on the present model.

\section{Data Availability}

\label{data}
This section provides the link to the codes and data, which includes the code and input file for all the examples in the article after the acceptance of the article.



\appendix

\bibliographystyle{elsarticle-num}
%
%
\bibliography{references}

\newpage

\section{:Stress Distribution in Double Notched tension Specimens}

\begin{figure}[H]
\begin{center}
\includegraphics[width=0.60\linewidth]{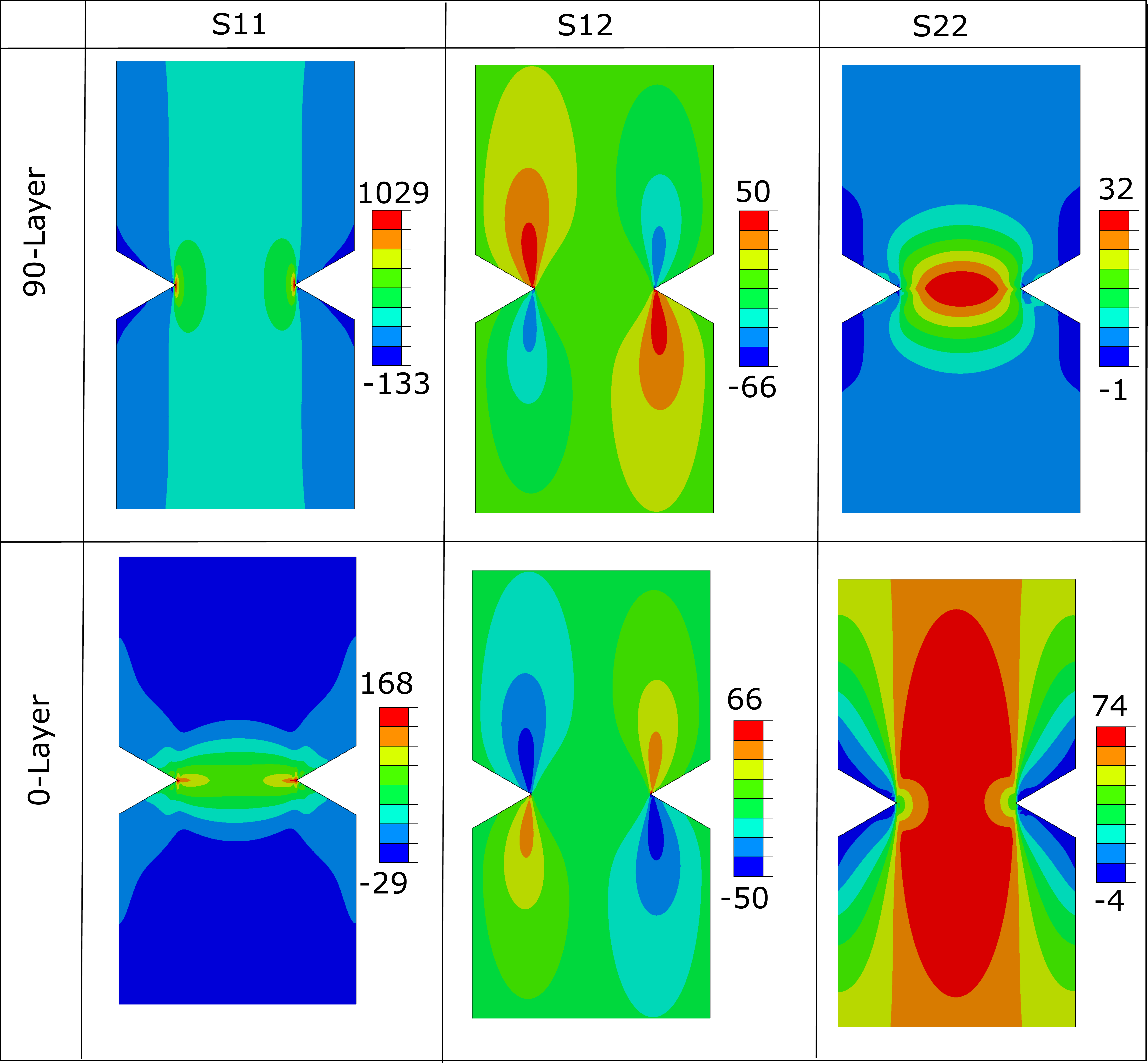}
\caption{Stress distribution in each layer of laminate $[0/90]_s$. All the stress are in MPa. }
\label{fig:ctlam_pf_stress}
\end{center}
\end{figure}

\begin{figure}[H]
\begin{center}
\includegraphics[width=0.60\linewidth]{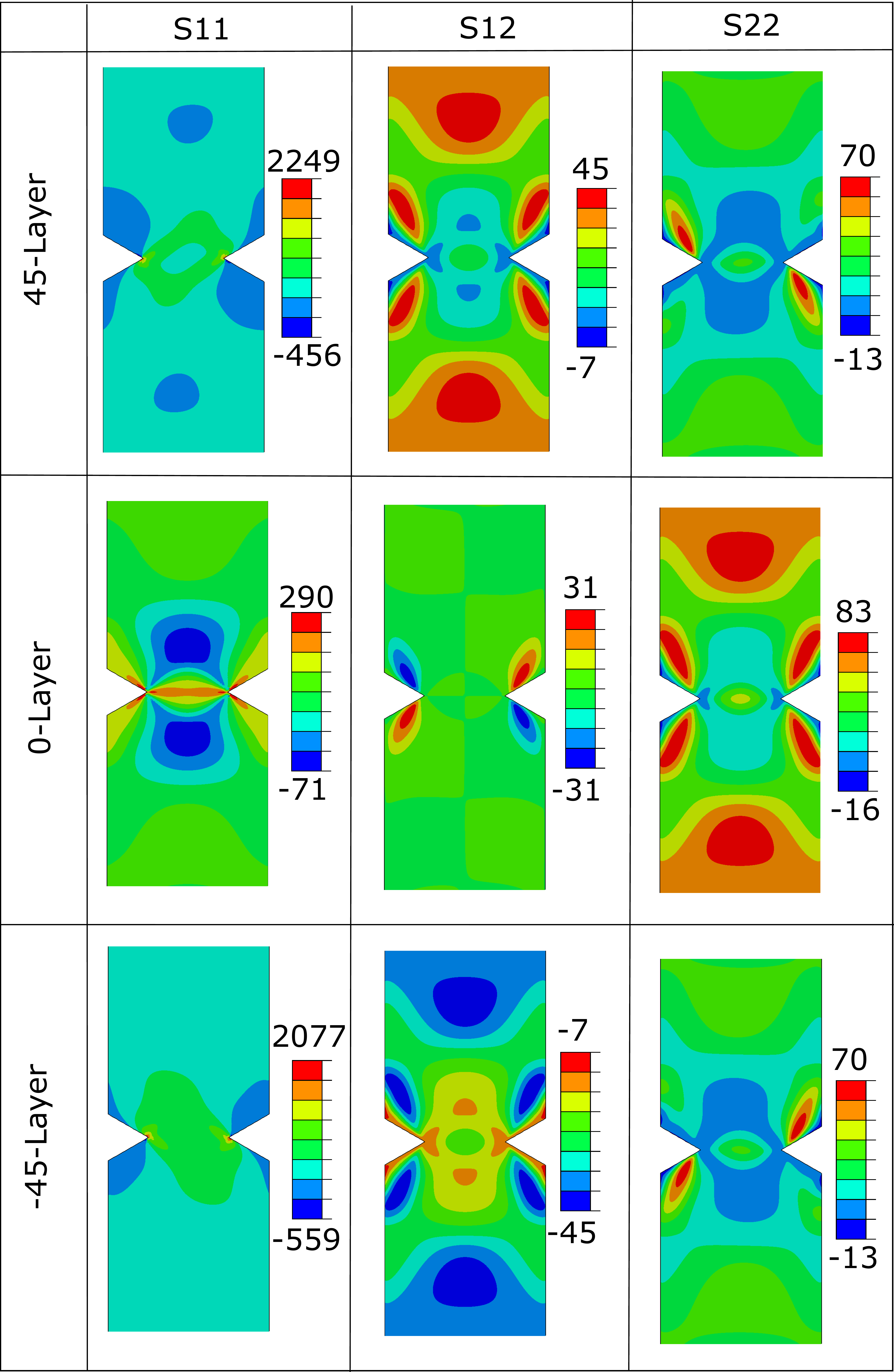}
\caption{Stress distribution in each layer of the laminate $[45/90/-45]_s$. All the stresses are in MPa. }
\label{DNstress2}
\end{center}
\end{figure}

\end{document}